\documentclass[amsmath,amssymb,aps,floatfix,pre,showpacs,superscriptaddress,twocolumn]{revtex4}

\usepackage{graphicx}
\usepackage{bm}
\usepackage{color} 

\begin{document}


\title{Stochastic calculus for uncoupled continuous-time random walks}

\author{Guido Germano}
\email{guido.germano@staff.uni-marburg.de}
\homepage{www.staff.uni-marburg.de/~germano}
\affiliation{Fachbereich Chemie und WZMW, Philipps-Universit\"at Marburg,
35032 Marburg, Germany}

\author{Mauro Politi}
\email{mauro.politi@unimi.it}
\affiliation{Fachbereich Chemie und WZMW, Philipps-Universit\"at Marburg,
35032 Marburg, Germany}
\affiliation{Dipartimento di Fisica, Universit\`a degli Studi di Milano,
Via Celoria 16, 20133 Milano, Italy}

\author{Enrico Scalas}
\email{enrico.scalas@mfn.unipmn.it}
\homepage{www.mfn.unipmn.it/~scalas}
\affiliation{Dipartimento di Scienze e Tecnologie Avanzate, Universit\`a del
Piemonte Orientale ``Amedeo Avogadro'', Via Vincenzo Bellini 25 G, 15100
Alessandria, Italy}

\author{Ren\'e L.\ Schilling}
\email{rene.schilling@tu-dresden.de}
\homepage{www.math.tu-dresden.de/sto/schilling}
\affiliation{Institut f\"ur Mathematische Stochastik, Technische Universit\"at
Dresden, 01062 Dresden, Germany}

\date{\today}



\begin{abstract}
The continuous-time random walk (CTRW) is a pure-jump stochastic process with
several applications in physics, but also in insurance, finance and economics.
A definition is given for a class of stochastic integrals driven by a CTRW,
that includes the It\=o and Stratonovich cases. An uncoupled CTRW with
zero-mean jumps is a martingale. It is proved that, as a consequence of the
martingale transform theorem, if the CTRW is a martingale, the It\=o integral
is a martingale too. It is shown how the definition of the stochastic integrals
can be used to easily compute them by Monte Carlo simulation. The relations
between a CTRW, its quadratic variation, its Stratonovich integral and its
It\=o integral are highlighted by numerical calculations when the jumps in
space of the CTRW have a symmetric L\'evy $\alpha$-stable distribution and its
waiting times have a one-parameter Mittag-Leffler distribution. Remarkably
these distributions have fat tails and an unbounded quadratic variation.
In the diffusive limit of vanishing scale parameters, the probability density
of this kind of CTRW satisfies the space-time fractional diffusion equation
(FDE) or more in general the fractional Fokker-Planck equation, that generalize
the standard diffusion equation solved by the probability density of the Wiener
process, and thus provides a phenomenologic model of anomalous diffusion.
We also provide an analytic expression for the quadratic variation of the
stochastic process described by the FDE, and check it by Monte Carlo.
\end{abstract}

\maketitle

\section{Introduction}\label{sec:introduction}

\subsection{The continuous-time random walk}\label{sec:ctrw}

The continuous-time random walk (CTRW) is a pure-jump stochastic process used
as a model for standard and anomalous diffusion when the sojourn time at a site
is much greater than the time needed to jump to a new position, i.e.\ when
jumps can be considered instantaneous events. The CTRW has been introduced in
physics by Montroll and Weiss \cite{montroll65}; other seminal papers on its
application to standard and anomalous transport phenomena are due to Scher and
Lax \cite{scher73a,scher73b} and to Montroll and Scher \cite{montroll73,
scher75}. More recently, Shlesinger wrote a review that contributed to further
popularize the CTRW \cite{shlesinger96}; theoretical, numerical, and empirical
studies on the CTRW have been discussed by Weiss \cite{weiss94}, Metzler and
Klafter \cite{metzler00,metzler04}, and some authors of the present paper
\cite{scalas06,fulger08}.

In a CTRW, if $X(t)$ denotes the position of a diffusing particle at time $t$,
$\xi_i = X(t_i) - X(t_{i-1})$ denotes a random jump occurring at a random time
$t_i$, and $\tau_i = t_i - t_{i-1}$ is the waiting or sojourn or interarrival
or duration time between two consecutive jumps, one has
\begin{equation}\label{ctrw}
X(t) \stackrel{\text{def}}{=} S_{N(t)}
\stackrel{\text{def}}{=} \sum_{i=1}^{N(t)} \xi_i,
\end{equation}
where $t_0 = 0$, $X(0) = 0$ and $N(t)$ is a counting random process that gives
the number of jumps up to time $t$. Throughout this paper, we assume that
\begin{itemize}
\item[-] the jumps $\xi_i,\ i=1,2,\ldots$ are independent and identically
distributed (iid) random vectors in $\mathbb{R}^d$, $d=1,2,\ldots$
\cite{meerschaert01};
\item[-] the waiting times $\tau_i,\ i=1,2,\ldots$ are iid random variables in
$\mathbb{R}_+$;
\item[-] the families $(\xi_i,\ i=1,2,\ldots)$ and $(\tau_i,\ i=1,2,\ldots)$
are independent.
\end{itemize}
The third assumption means that we consider a so-called uncoupled CTRW. The
first two assumptions entail that the joint distribution of any pair $(\xi_i,
\tau_i)$ does not depend on $i$. If, in the uncoupled case, the law of $(\xi_i,
\tau_i)$ is given by a density function $\varphi(\xi,\tau)$, the independence
of $\xi_i$ and $\tau_i$ means that it can be factorized in terms of the
marginal probability densities for jumps $\lambda(\xi)$ and waiting times
$\psi(\tau)$: $\varphi(\xi,\tau) = \lambda(\xi)\psi(\tau)$.

\begin{figure}[h]
\includegraphics[angle=-90,width=\columnwidth]{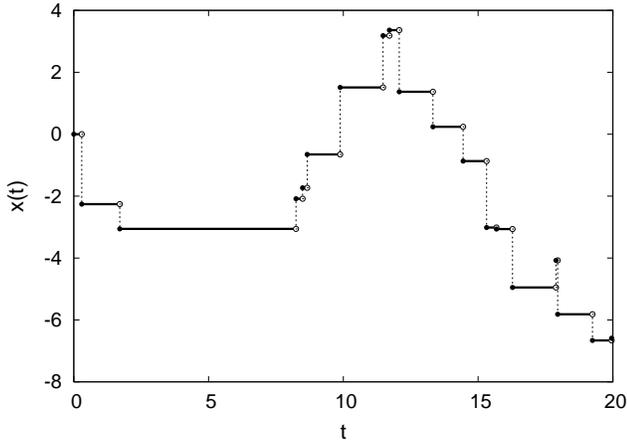}
\caption{\label{fig:walks}
Realization of a CTRW with exponentially distributed waiting times ($\gamma_t =
1$) and standard normally distributed jumps ($\mu = 0$ and $\sigma=1$).}
\end{figure}

Eq.~(\ref{ctrw}) means that a CTRW is a random sum of independent random
variables. The process of the jump times
\begin{equation}\label{renewal}
t_n = \sum_{i=1}^n \tau_i, \,\, t_0 = 0,
\end{equation}
is a renewal point process. Therefore, a CTRW can be seen as a compound renewal
process \cite{cox67,feller71,cox79}. The existence of an uncoupled CTRW can be
proved, based on the corresponding theorems of existence for renewal processes
and discrete-time random walks \cite{billingsley79}. C\`adl\`ag
(right-continuous with left limit) realizations of a CTRW can be easily and
exactly generated by Monte Carlo simulation and plotted \cite{fulger08}.
This is illustrated in Fig.~\ref{fig:walks}.
An uncoupled CTRW is Markovian if and only if the waiting time distribution is
exponential, i.e.\ $\psi(\tau) = \exp(-\tau/\gamma_t)/\gamma_t$ \cite{hoel72,
cinlar75}. An uncoupled CTRW belongs to the class of semi-Markov processes
\cite{cinlar75,flomenbom05,flomenbom07,janssen07}, i.e.\ for any $A \subset
\mathbb{R}^d$ and $t > 0$ we have
\begin{multline}\label{semi-markov}
P(S_n \in A, \tau_n \leq t \,|\, S_0, \ldots, S_{n-1}, \tau_1, \ldots,
\tau_{n-1}) \\
= P(S_n \in A, \tau_n \leq t \,|\, S_{n-1})
\end{multline}
and, if we fix the position $S_{n-1} = y$ of the diffusing particle at time
$t_{n-1}$, the probability on the right will be independent of $n$. In the
generic coupled case, if the law of $(\xi_n,\tau_n)$ is given by a density
function $\varphi(\xi,\tau)$, we can use $S_n = S_{n-1} + \xi_n$ and rewrite
this as
\begin{equation}\label{semi-markov2}
P(S_n \in A, \tau_n \leq t \,|\, S_{n-1})
= \int_A \int_0^t \varphi(x-S_{n-1},\tau) \, d\tau dx.
\end{equation}
This can be shown as follows. Let $I_A(x)$ denote the indicator function that
yields 1 if $x \in A$ and 0 otherwise. Probabilities can be replaced with
expectations writing $P(x \in A) = E[I_A(x)]$. Moreover, one has $I_A I_B =
I_{A \cap B}$. Thus, if $B = (0,t]$:
\begin{multline}\label{semi-markov-proof}
P(S_n \in A, \tau_n \in B \,|\, S_{n-1}) \\
\begin{aligned}
&= E[I_A(S_n) I_B(\tau_n) \,|\, S_{n-1}] \\
&= E[I_A(S_{n-1}+\xi_n) I_B(\tau_n) \,|\, S_{n-1}] \\
&= \int_{\mathbb{R}^d} \int_0^\infty I_A(S_{n-1}+\xi) I_B(\tau)
   \varphi(\xi,\tau) \, d\tau d\xi \\
&= \int_{\mathbb{R}^d} \int_B I_A(S_{n-1}+\xi) \varphi(\xi,\tau)\,d\tau d\xi \\
&= \int_{\mathbb{R}^d} \int_0^t I_A(x) \varphi(x-S_{n-1},\tau) \, d\tau dx \\
&= \int_A \int_0^t \varphi(x-S_{n-1},\tau) \, d\tau dx.
\end{aligned}
\end{multline}

Montroll and Weiss wrote Eq.~(\ref{semi-markov2}) as an integral equation for
the probability density $p_X(x,t)$ of finding the particle in position $x$ at
time $t$ in terms of the joint probability density $\varphi(\xi,\tau)$ of the
jumps $\xi$ and waiting times $\tau$:
\begin{equation}\label{integralequation}
p_X(x,t) = \delta(x) \Psi(t) + \int_{\mathbb R^d} \int_0^t \varphi(\xi,\tau)
p_X(x-\xi,t-\tau) \, d\tau d\xi,
\end{equation}
where $\Psi(t) = 1 - \int_0^t \psi(\tau) \, d\tau$ is the complementary
cumulative distribution function for the waiting times, also called survival
function. This can be shown observing that
\begin{equation}\label{0star}
P(X(t) \in dx \,|\, X(0) = 0) = p_X(x,t)\,dx
\end{equation}
and
\begin{multline}\label{1star}
P(X(t) \in dx \,|\, X(t') = x') \\
\begin{aligned}
&= P(X(t-t') \in dx \,|\, X(0) = x') \\
&= P(X(t-t') - x' \in dx \,|\, X(0) = 0) \\
&= p_X(x-x',t-t')\,dx
\end{aligned}
\end{multline}
because the increments in time and space are iid and hence homogeneous.
Moreover, from Eq.~(\ref{semi-markov2}),
\begin{gather}\label{2star}
P(S_1 \in dx, \tau_1 \in dt \,|\, S_0 = 0) = \varphi(x,t)\,dxdt.
\end{gather}
The probability in Eq.~(\ref{0star}) can be decomposed depending on the
duration of the first jump $\tau_1$ with respect to $t$:
\begin{multline}
P(X(t) \in dx \,|\, X(0) = 0) \\
\begin{aligned}
&= \phantom{+} P(X(t) \in dx, \tau_1 > t \,|\, X(0) = 0)\, \\
&\phantom{=} + P(X(t) \in dx, \tau_1 \leq t \,|\, X(0) = 0).
\end{aligned}
\end{multline}
The part without a jump before $t$ is given by
\begin{align}\label{case1}
P(X(t) \in dx, \tau_1 > t \,|\, X(0)=0) &= P(\tau_1>t)\delta(x)\,dx \nonumber\\
                                        &= \delta(x) \Psi(t) \, dx.
\end{align}
The other part is given by
\begin{multline}\label{case2}
P(X(t)\in dx, \tau_1\leq t \,|\, X(0)=0) \\
\begin{aligned}
&\stackrel{\phantom{\text{Eq.~(9)}}}{=} \int_{\mathbb{R}^d}\int_0^t
P(X(t)\in dx \,|\, X(t')=x') \\
&\qquad\qquad\qquad\times P(S_1\in dx', \tau_1\in dt' \,|\, S_0=0) \\
&\stackrel{\text{Eq.~(\ref{2star})}}{=} \int_{\mathbb{R}^d}\int_0^t
P(X(t)\in dx \,|\, X(t')=x')\,\varphi(x',t')\,dt'dx' \\
&\stackrel{\text{Eq.~(\ref{1star})}}{=} \int_{\mathbb{R}^d}\int_0^t
p_X(x-x',t-t')dx\,\varphi(x',t')\,dt'dx'\\
&\stackrel{\phantom{\text{Eq.~(9)}}}{=} \left[\int_{\mathbb{R}^d}\int_0^t
\varphi(\xi,\tau)\,p_X(x-\xi,t-\tau)\,d\tau d\xi \right]\,dx.
\end{aligned}
\end{multline}
Combining Eqs.~(\ref{case1}) and (\ref{case2}) yields
Eq.~(\ref{integralequation}). Notice that the latter just gives a one-point
probability density, which is not enough to characterize a stochastic process
without further assumptions.

Eq.~(\ref{integralequation}) can be solved in the Fourier-Laplace domain,
\begin{equation}
\widehat{\widetilde{p}}(k,s) = \frac{1}{1-\widehat{\widetilde{\varphi}}(k,s)}\,
\frac{1 - \widetilde{\psi}(s)}{s},
\end{equation}
where the Fourier and Laplace transforms are defined as
\begin{gather}
\widehat{f}(k) = \mathcal{F}_x[f(x)](k) =
\int_{-\infty}^{+\infty}f(x) e^{i k x}\,dx,\quad k \in \mathbb{R},\\
\widetilde{f}(s) = \mathcal{L}_t[f(t)](s) = \int_0^\infty f(t) e^{-st}\,dt,
\quad s \in \mathbb{C}.
\end{gather}
The inverse transforms to the space-time domain are possible in the uncoupled
case, i.e.\ when $\varphi(\xi,\tau) = \lambda(\xi)\psi(\tau)$; this leads to a
series expression written in terms of the probability $P(N(t)=n) = p_N(n,t)$ of
the counting process $N(t)$, and the $n$-fold convolution $\lambda^{*n}(x)$ of
the marginal probability density of jumps $\lambda(\xi)$:
\begin{equation}\label{MWsolution}
p_X(x,t) = \sum_{n=0}^\infty p_N(n,t) \lambda^{*n}(x).
\end{equation}
The method using integral transforms is described in several papers, including
the original one by Montroll and Weiss.
However, Eq.~(\ref{MWsolution}) can also be derived directly by probabilistic
considerations. Indeed, Eq.~(\ref{ctrw}) is a random sum of iid random
variables. This means that any position $x$ can be reached at time $t$ by a
finite number $n$ of jumps. The probability of reaching position $x$ at time
$t$ in exactly $n$ jumps is $p_N(n,t) \lambda^{*n}(x)$. Eq.~(\ref{MWsolution})
follows given that these events are mutually exclusive. Note that $p_N(0,t)
\lambda^{*0}(x)$ coincides with the singular term $\delta(x) \Psi(t)$, meaning
that the distribution function for $x$ has a jump at position $x=0$ of height
$\Psi(t)$.

A CTRW with exponential waiting times is called a compound Poisson process
(CPP), as in this case
\begin{equation}
p_N(n,t;\gamma_t) = \exp(-t/\gamma_t)\frac{(t/\gamma_t)^n}{n!}.
\end{equation}
A CPP is not only a Markov, but also a L\'evy process. This means that it has
independent and time-homogeneous (stationary) increments. In the L\'evy case
$p_X(x,t)$, even $p_X(x,1)$, fully characterizes the stochastic process defined
by Eq.~(\ref{ctrw}) \cite{billingsley79,bertoin96,sato99}; this is due to the
infinite divisibility and the fact that the increments are stationary and
independent. For a normal CPP, i.e.\ a CPP with normally distributed jumps,
the $n$-fold convolution $\lambda^{*n}(x)$ of $N(\mu,\sigma^2)$ can be evaluated
as $N(n\mu,n\sigma^2)$, leading to
\begin{multline}\label{NCPPdens}
p_X(x,t;\mu,\sigma,\gamma_t) = \exp(-t/\gamma_t) \\
\times \sum_{n=0}^\infty \frac{(t/\gamma_t)^n}{n!}\frac{1}{\sqrt{2\pi n}\sigma}
\exp \left( -\frac{(x-n\mu)^2}{2n\sigma^2} \right).
\end{multline}

\subsection{The CTRW in physics, insurance, finance, and economics}

Since the seminal paper by Montroll and Weiss \cite{montroll65}, there has been
much scientific activity on the application of the CTRW to important physical
problems. One line of research investigated anomalous relaxation related to
power-law tails of the waiting time distribution as well as the asymptotic
behaviour of the CTRW for large times \cite{montroll73,shlesinger74,tunaley74,
tunaley75,tunaley76,shlesinger82}. As mentioned above, Klafter and Metzler have
extensively reviewed these and subsequent studies \cite{metzler00,metzler04}.
Furthermore, in their book, ben-Avraham and Havlin have discussed the
applications to physical chemistry \cite{benavraham00}. Here, it is worth
mentioning the recent work on the relation between the CTRW and fractional
diffusion that can be traced to papers by Balakrishnan and Hilfer
\cite{balakrishnan85,hilfer95} and has been thoroughly discussed in Refs.\
\cite{scalas04,scalas06,fulger08}. Some specific applications include, e.g.,
plasmas \cite{negrete05} and biopolymers \cite{dubbeldam07a,dubbeldam07b}.

The CTRW has been applied also in insurance, finance, and economics. Even if
well-known in the field of econophysics \cite{scalas06,masoliver06}, these
applications deserve a short summary.

In ruin theory for insurance companies, the jumps $\xi_i$ are interpreted as
claims and they are positive random variables; $t_i$ is the instant at which
the $i$-th claim is paid \cite{embrechts97}.

In mathematical finance, if $P_A (t)$ is the price of an asset at time $t$
and $P_A (0)$ is the price of the same asset at a previous reference time
$t_0 = 0$, then $X(t) = \log(P_A (t)/P_A (0))$ represents the log-return (or
log-price) at time $t$. In regulated markets using a continuous double-auction
trading mechanism, such as stock markets, prices vary at random times $t_i$,
when a trade takes place, and $\xi_i = X(t_i) - X(t_{i-1}) = \log(P_A (t_i)/P_A
(t_{i-1}))$ is the tick-by-tick log-return, whereas $\tau_i = t_i - t_{i-1}$ is
the intertrade duration; for more details, see \cite{scalas06,masoliver06,
cartea07} and references contained therein.

In the theory of economic growth, $\xi_i$ represents a growth shock, which can
actually be both positive and negative, $X(t)$ is the logarithm of a firm's
size or of an individual's wealth, and $\tau_i$ is the time interval between
two consecutive growth shocks; see \cite{scalas06} and references therein.

\subsection{Motivation for the study of stochastic integrals driven by a CTRW
and link with fractional calculus}
\label{sec:motivation}

Given the wide range of applications of the CTRW overviewed in the previous
subsection, it is relevant to study diffusive stochastic differential equations
whose driving noise is defined in terms of a CTRW:
\begin{equation}\label{sde}
dZ = a(Z,t) dt + b(Z,t) dX.
\end{equation}
Here $Z(X,t)$ is the unknown random function, $a(Z,t)$ and $b(Z,t)$ are known
functions of $Z$ and time $t$, and $dX$ represents the CTRW `measure' with
respect to which stochastic integrals are defined. In order to give a rigorous
meaning to such an expression, some constraints on the properties of the CTRW
are necessary. In a recent paper, the theory has been discussed for stochastic
integration on a time-homogeneous (stationary) CTRW --- i.e., the already
mentioned CPPs \cite{zygadlo03}. Although the theory reported there was already
well known by mathematicians and has been used in finance for option pricing
since 1976 \cite{merton76}, that paper contains useful material and is written
in a way that is clear and appealing for physicists. Here, inspired by
Ref.~\cite{zygadlo03}, the theory will be further discussed and developed.

Consider a CTRW $X(t)$ whose jumps in space $\xi_i$ are distributed according
to the symmetric L\'evy $\alpha$-stable law, $\alpha \in (0,2]$, whose density
can be expressed as a series or, more conveniently, as the inverse Fourier
transform of its characteristic function: 
\begin{equation}\label{levy}
L_\alpha(\xi;\gamma_x)
= \mathcal{F}^{-1}_k\left[ \exp\left(-|\gamma_x k|^\alpha\right) \right](\xi).
\end{equation}
For $\alpha = 2$ this corresponds to a Gaussian with standard deviation
$\sigma = \sqrt{2}\gamma_x$. Let the waiting times $\tau_i$ of the CTRW have
the probability density
\begin{equation}\label{mldensity}
\psi_\beta(\tau;\gamma_t)
= -\frac{d}{d\tau}E_\beta\left(-(\tau/\gamma_t)^\beta\right),
\end{equation}
where $E_\beta(z),\ \beta \in (0,1],$ is the one-parameter Mittag-Leffler
function \cite{gorenflo02,podlubny05,hilfer06}:
\begin{equation}
E_\beta(z) = \sum_{n=0}^\infty \frac{z^n}{\Gamma(\beta n+1)}, \quad
z \in \mathbb{C}.
\end{equation}
For a real argument $z = t \in \mathbb{R}$ and $\beta = 1$ this corresponds to
an exponential function. When $\beta < 1$, $E_\beta(-t^\beta)$ is approximated
for small values of $t$ by a stretched exponential decay (Weibull function),
$\exp\left(-t^\beta/\Gamma(1+\beta)\right)$, and for large values of $t$ by a
power law, $t^{-\beta}/\Gamma(1-\beta)$.

In the diffusive limit for $X(t)$, when the scale parameters $\gamma_x$ of
the jumps and $\gamma_t$ of the waiting times vanish satisfying the scaling
relation $\gamma_x^\alpha/\gamma_t^\beta = D$, if in Eq.~(\ref{sde}) $a = 0$
and $b = 1$ the probability density $p_Z(z,t) = p_X(x,t;\gamma_x,\gamma_t)$
converges to the solution of the space-time fractional diffusion equation (FDE)
\cite{samko93,podlubny99}
\begin{gather}\label{fracdiff}
\frac{\partial^\beta}{\partial t^\beta} u_X(x,t;D) = D
\frac{\partial^\alpha}{\partial |x|^\alpha} u_X(x,t;D)\\ \nonumber
u_X(x,0^+;D) = \delta (x), \quad x \in \mathbb{R}, \quad t \in \mathbb{R}_+.
\end{gather}
The space-fractional derivative of order $\alpha \in (0,2]$ is defined
according to Riesz:
\begin{equation}
\frac{d^\alpha}{d|x|^\alpha} f(x)
= \mathcal{F}^{-1}_k\left[-|k|^\alpha \widehat{f}(k)\right](x).
\end{equation}
The time-fractional derivative of order $\beta \in (0,1]$ is defined in the
sense of Caputo
\begin{equation}
\frac{d^\beta}{dt^\beta}f(t)
= \mathcal{L}^{-1}_s\left[s^\beta\widetilde{f}(s)-s^{\beta-1}f(0^+)\right](t).
\end{equation}
The FDE is a generalization of the standard diffusion equation, that results
for $\alpha = 2$ and $\beta = 1$; in this case the solution $u_X(x,t;D)$ of the
Cauchy problem given by Eq.~(\ref{fracdiff}) is the one-point probability
density of the Bachelier-Wiener process or Brownian motion $B(t)$,
\begin{equation}\label{wiener}
u_X(x,t;D) = \frac{1}{\sqrt{4\pi Dt}}\exp\left(-\frac{x^2}{4Dt}\right),
\end{equation}
and $X(t)$ is the NCPP introduced at the end of Sec.~\ref{sec:ctrw}.
The general solution of the FDE was worked out in the Fourier-Laplace domain:
\begin{equation}
\widehat{\widetilde{u}}_X(k,s) = \frac{s^{\beta-1}}{D|k|^\alpha+s^\beta}.
\end{equation}
Because
\begin{equation}
\mathcal{L}^{-1}_s\left[\frac{s^{\beta-1}}{D|k|^\alpha+s^\beta}\right](t)
= E_\beta(-D|k|^\alpha t^\beta)
\end{equation}
defining $\kappa = kt^{\beta/\alpha}$ and the time-independent Green function
\begin{equation}\label{scalingfunction}
G_{\alpha,\beta}(\xi;D) = \mathcal{F}^{-1}_\kappa
\big[E_\beta(-D|\kappa|^\alpha)\big](\xi),
\end{equation}
the solution of the FDE, Eq.~(\ref{fracdiff}), can be expressed in the
space-time domain as
\begin{equation}\label{greenfunction}
u_X(x,t;D) = t^{-\beta/\alpha}\,G_{\alpha,\beta}(xt^{-\beta/\alpha};D).
\end{equation}
These results are a consequence of a generalized central limit theorem for
sequences of random variables \cite{scalas04}. A simpler derivation can be
found in Ref.~\cite{scalas06}. For computational details see
Sec.~\ref{sec:simulation} and Ref.~\cite{fulger08}.
If $a(x,t)$ and $b(x,t)$ are not constant, a fractional Fokker-Planck equation
for $u_X(x,t;D)$ has been proposed in the diffusive limit \cite{metzler99a,
metzler99b,metzler99c,barkai00,metzler00,magdziarz07} starting from a
generalized master equation \cite{metzler99c} or a CTRW \cite{barkai00}.
For the NCPP this reduces to the standard Fokker-Planck equation
\cite{vankampen81,risken92}.

Without taking the diffusive limit, and if $a = 0$ and $b = 1$, the time
evolution of the probability density $p_X(x,t)$ is given by the Montroll-Weiss
integral equation (\ref{integralequation}). The uncoupled case of the latter
can be presented alternatively in an integro-differential form
\cite{mainardi00},
\begin{multline}\label{integrodiffeq}
\int_0^t \Phi(t-\tau) \frac{\partial}{\partial \tau} p_X(x,\tau) \, d\tau \\
= -p_X(x,t) + \int_{-\infty}^{+\infty} \lambda(x-\xi)p_X(\xi,t) \, d\xi,
\end{multline}
that can be interpreted as a time evolution equation of Fokker-Planck type.
It involves the time derivative of $p_X(x,t)$ and an auxiliary function
$\Phi(t)$ defined through its Laplace transform as $\widetilde{\Phi}(s) =
\widetilde{\Psi}(s)/\widetilde{\psi}(s)$, so that $\Psi(t) = \int_0^t
\Phi(t-\tau) \psi(\tau) \, d\tau$. This approach has been generalized studying
scores of possible kinetic equations for non-Markovian processes \cite{mura08}.
What follows in the next sections is valid
without necessarily taking the diffusive limit. Nevertheless, the latter
is important because it motivates our particular choice for the marginal
distributions of jumps and waiting times, and because it provides analytic
expressions that can be compared to our Monte Carlo results as shown in
Sec.~\ref{sec:simulation}.

\section{Stochastic integrals}\label{sec:theory}

In Ref.~\cite{zygadlo03}, the stochastic integral is never explicitly defined.
However, starting from the fact that sample paths of a CTRW can be represented
by step functions, it is possible to give an explicit formula.

\subsection{Definitions}

Some heuristic manipulations are useful for the definition of the stochastic
integral
\begin{equation}\label{integral}
J(t) = \int_0^t Y(s) \, dX(s),
\end{equation}
where $X(t)$ and $Y(t)$ are synchronous CTRWs, i.e.\ their jumps happen at the
same times $t_i,\ i = 1,\ldots,N(t)$. Though an interesting case is often $Y(t)
= G(X(t))$ with a suitable function $G(X)$, the jumps of $Y(t)$ and $X(t)$ at
$t = t_i$ may be independent as well. Eq.~(\ref{ctrw}) defining $X(t)$ can be
written in terms of the right-continuous variant of Heaviside's step function
$\theta(t)$, which is $0$ for $t < 0$ and $1$ for $t \geq 0$:
\begin{equation}\label{ctrwheaviside}
X(t) = \sum_{i=1}^{N(t)} \xi_i \theta(t-t_i).
\end{equation}
Using the fact that the `derivative' of Heaviside's $\theta$ function
$\theta(t-t_i)$ is Dirac's $\delta$ function $\delta(t-t_i)$, one can write
\begin{equation}\label{measure}
d X(t) = \sum_{i=1}^{N(t)} \xi_i \delta(t-t_i) \, dt,
\end{equation}
which means that $\Delta X(t_i) \stackrel{\text{def}}= X(t_i)-X(t_i^-) = \xi_i$
with $X(t_i^-) = \lim_{s \to t_i^-} X(s) 
\stackrel{\text{def}}= \lim_{s \to t_i, s < t_i} X(s)$. Note that $\delta(t)$
is not a proper function, but rather a distribution in the sense of Sobolev and
Schwartz \cite{gelfand64}. Writing Eq.~(\ref{measure}) with $t_i^-$ in place of
$t_i$, inserting it into Eq.~(\ref{integral}), and using the properties of
Dirac's $\delta$ function, we get the exact expression (no limit needed: recall
that the number of jumps $N(t)$ between $0$ and $t$ is a random finite integer)
\begin{eqnarray}\label{ito_integral}
I(t) &\stackrel{\text{def}}{=}& \int_0^t Y(s^-)\,dX(s)
=  \sum_{i=1}^{N(t)} Y(t_i^-) \xi_i \nonumber \\
&=& \sum_{i=1}^{N(t)} Y(t_i^-) (X(t_i) - X(t_i^-)).
\end{eqnarray}
The choice $Y(s^-)$ for the integrand makes $I(t)$ a martingale if $X(t)$ is a
martingale, as will be explained below. This
naive definition works nicely if the driving noise is a step function with jump
times $t_i$ and jumps $\xi_i = X(t_i) - X(t_i^-)$; if $Y(t)$ and $X(t)$ jump at
the same time we even have $Y(t_i^-) = Y(t_{i-1})$. As soon as one wants to go
beyond this situation, measurability and convergence become an issue.
This observation prompted K.~It\=o to use martingale convergence theorems to
tackle the convergence for a large class of integrators \cite{protter04}. To do
so we must make sure that $I(t)$ is a martingale whenever $X(t)$ is. For this
we assume that $Y(t)$ is adapted i.e.\ measurable with respect to the natural
filtration generated by the driving noise: $\mathcal{F}_t = \sigma(X(s) : s\leq
t)$. Therefore the integrand $Y(t_i^-)$ in Eq.~(\ref{ito_integral}) becomes
statistically independent of the increment $\xi_i = X(t_i) - X(t_i^-)$ and we
end up with a stochastic integral $I(t)$ that is a martingale; see the next
section for details. The fact that we evaluate $Y(t)$ at the left end-point
$t_i^-$ of the `infinitesimal interval' $[t_i^-,t_i]$ makes the integrand
non-anticipating and adapted, i.e.\ independent of the increment. This can be
seen as a causality requirement: one does not want $Y(t)$ to anticipate the
future behavior of $\xi(t)$ \cite{paul00}. An elementary introduction to the
concept of a non-anticipating function can be found in Ref.~\cite{gardiner85}.
Any adapted process with right-continuous (or left-continuous) paths is
progressively measurable.

In Eq.~(\ref{ito_integral}) we might equally well choose to evaluate $Y(t)$ in
the right end-point $t_i$ of the infinitesimal interval $[t_i^-,t_i]$,
corresponding to the right-continuous variant of Heaviside's $\theta$ function
in Eq.~(\ref{ctrwheaviside}), or in any intermediate point $t_i^\vartheta$.
This means, however, loosing the martingale property of the stochastic
integral. The effects on the formulae for such a choice can be nicely described
for random step functions $Y(t)$ and $X(t)$ jumping at the same times $t_1,
t_2, \ldots, t_n$. Write
\begin{eqnarray}\label{stoch_integral}
J_\vartheta (t) &\stackrel{\text{def}}{=}& \int_0^t Y(s_\vartheta) \, d X(s)
=  \sum_{i=1}^{N(t)} Y(t_i^\vartheta) \xi_i \\ \nonumber &=&
\sum_{i=1}^{N(t)} [(1-\vartheta)Y(t_i^-) + \vartheta Y(t_i)] [X(t_i)-X(t_i^-)]
\end{eqnarray}
for a parameter $\vartheta \in [0,1]$ that interpolates linearly between
$Y(t_i^-) = Y(t_{i-1})$ and $Y(t_i)$, resulting in a continuous class of
stochastic integrals. The choice $\vartheta = 0$ gives the It\=o integral
$J_0(t) = I(t)$. For any value of $\vartheta$ the integral is
a right-continuous function with jump $\Delta J_\vartheta(t_i)
= J_\vartheta(t_i) - J_\vartheta(t_i^-) = Y(t_i^\vartheta) \Delta X(t_i)$.

Eq.~(\ref{stoch_integral}) can be rearranged to
\begin{equation}\label{compensator}
J_\vartheta (t) = J_{1/2}(t) + \left( \vartheta - \frac{1}{2} \right) [X,Y](t),
\end{equation}
where
\begin{equation}\label{covariation}
[X,Y](t) \stackrel{\text{def}}{=}
\sum_{i=1}^{N(t)} [X(t_i)-X(t_i^-)][Y(t_i)-Y(t_i^-)]
\end{equation}
is the covariation or cross variation of $X(s)$ and $Y(s)$ for $s \in [0,t]$.
When $Y(s) = X(s)$, the quadratic variation $[X,X](t)$ is denoted simply as
$[X](t)$. Thus each member of the family of stochastic integrals with
$\vartheta \in [0,1]$ can be obtained adding a compensator to the Stratonovich
integral $J_{1/2}(t) = S(t)$. The latter corresponds to the symmetric variant
of Heaviside's step function, $\theta(t) = (\mathrm{sgn}\,t + 1)/2$, and is
particularly appealing because it can be computed according to the usual rules
of calculus. However, the It\=o integral has the advantage of being a
martingale, as proved in the next subsection. The distinction between integrals
with different values of $\vartheta$ disappears in the continuous limit for
processes with finite variation, e.g.\ continuously differentiable functions,
because this implies that their quadratic variation is zero \cite{protter04}.
Unless stated otherwise $\int Y(s) \, d X(s)$ denotes the It\=o integral,
while the Stratonovich integral is often indicated as $\int Y(s) \circ d X(s)$.

\subsection{Martingale property of the It\=o integral}

Although it is easy to simulate directly the stochastic process defined in
Eq.~(\ref{ito_integral}) --- see the next section for numerical examples ---
it is not so easy to derive its properties. Each term in the sum depends on the
previous ones and the nice properties of convolutions are not helpful here.
However, using the \emph{martingale transform theorem}, it is possible to
obtain conditions under which $I(t)$ is a martingale.

In order to define martingales, we need a filtered probability space $(\Omega,
\mathcal{F},(\mathcal{F}_t)_{t \geq 0}, P)$, where $(\mathcal{F}_t)_{t\geq 0}$
is a filtration --- i.e., an increasing family of sub $\sigma$-algebras ---
representing the information available up to time $t$. A martingale is a
stochastic process $X(t)$ for which the expected value $\mathbb{E} [|X(t)|]$
exists for $t \geq 0$ and the conditional expectation $\mathbb{E} [X(t)\,|\,
\mathcal{F}_s]$ is $X(s)$ for all $t \geq s$ \cite{williams91,protter04,
schilling05}.

Let us consider the natural filtration, that is the $\sigma$-algebra generated
by the CTRW itself: $\mathcal{F}_t = \sigma(x(s) : s \leq t) = \sigma(\xi_1,
\ldots, \xi_k; \tau_1, \ldots, \tau_k : k \leq N(t)) \stackrel{\text{def}}{=}
\mathcal{G}_{N(t)}$. Then $X(t)$ is a martingale with respect to
$\mathcal{F}_t$ if and only if the mean of the jumps $\mathbb{E}[\xi_i]$ is
zero. Denote by $(t_i,\xi_i)$ the time and height of the finitely many jumps
$i = N(s)+1, \ldots, N(t)$ occurring between $s$ and $t > s$. Then
\begin{equation}\label{ctrwmartingale1}
\mathbb{E}[X(t)\,|\,\mathcal{F}_s]
= X(s) + \sum_{i=N(s)+1}^{N(t)} \mathbb{E} [\xi_i\,|\,\mathcal{F}_s].
\end{equation}
Using the semi-Markov property, Eq.~(\ref{semi-markov}), we get for $i > N(s)$
\begin{equation}\label{independence}
\mathbb{E}[\xi_i\,|\,\mathcal{F}_s] = \mathbb{E}[\xi_i\,|\,\mathcal{G}_{N(s)}]
= \mathbb{E}[\xi_i\,|\,\xi_{N(s)}] = \mathbb{E}[\xi_i] = 0,
\end{equation}
thanks to the independence of $\xi_i$ and $\xi_1, \ldots, \xi_{N(s)}$.
Eq.~(\ref{ctrwmartingale1}) becomes
\begin{equation}\label{ctrwmartingale2}
\mathbb{E}[X(t)\,|\,\mathcal{F}_s] = X(s),
\end{equation}
which shows that $(X(t))_{t\geq 0}$ is indeed a martingale with respect to its
natural filtration.

Note that our argument is valid for a general uncoupled CTRW. We do not need
the independence of the increments $X(t+\Delta t)-X(t)$ of the process $X(t)$
for non-overlapping intervals. Of course, if we have independent increments,
i.e.\ a compound Poisson process $X(t)$, the proof becomes easier.

Let us now investigate the integral defined in Eq.~(\ref{ito_integral}) for a
martingale CTRW $X(t)$. If there is an arbitrary but finite number of jumps
between $s$ and $t > s$, one has
\begin{equation}\label{mart1}
\mathbb{E} [I(t)\,|\,\mathcal{F}_s] = I(s) + \sum_{i=N(s)+1}^{N(t)} \mathbb{E}
[Y(t_i^-) \xi_i \,|\, \mathcal{G}_{N(s)}];
\end{equation}
now, one observes that $\xi_i = X(t_i) - X(t_{i-1})$ and that the random sum in
Eq.~(\ref{mart1}) becomes
\begin{multline}\label{mart2}
\sum_{i=N(s)+1}^{N(t)} \mathbb{E} [Y(t_i^-) \xi_i \,|\, \mathcal{G}_{N(s)}] \\
= \sum_{i=N(s)+1}^{N(t)} \mathbb{E} [Y(t_i^-) (X(t_i) - X(t_{i-1})) \,|\,
\mathcal{G}_{N(s)}].
\end{multline}
If $Y(t)$ is measurable with respect to $\mathcal{F}_t = \mathcal{G}_{N(t)}$,
then $Y(t_i^-)$ is $\mathcal{G}_{N(t_i^-)}$-measurable. Since $N(t_i^-) =
N(t_{i-1})$, this means that $Y(t_i^-)$ is $\mathcal{G}_{N(t_{i-1})} =
\mathcal{G}_{i-1}$-measurable; this is to say that $Y(t_i^-)$ is predictable
for the filtration $\mathcal{G}_i$, i.e.\ the value of $Y(t_i^-)$ is known at
time $t_{i-1}$. Whenever for each $i$ the expression $Y(t_i^-) (X(t_i) -
X(t_{i-1}))$ has a finite absolute mean --- e.g., if the process $Y(t_i^-)$ is
bounded --- we have
\begin{multline}\label{marttransf}
\mathbb{E} [Y(t_i^-) (X(t_i) - X(t_{i-1})) \,|\, \mathcal{G}_{N(s)}] \\
\begin{aligned}
&= \mathbb{E} \big[ \mathbb{E} [Y(t_i^-) (X(t_i) - X(t_{i-1})) \,|\,
\mathcal{G}_{i-1}] \,|\, \mathcal{G}_{N(s)}\big]\\
&= \mathbb{E} \big[ Y(t_i^-)\, \mathbb{E} [(X(t_i) - X(t_{i-1})) \,|\,
\mathcal{G}_{i-1}] \,|\, \mathcal{G}_{N(s)}\big]
\end{aligned}
\end{multline}
In the above calculation we have used the fact that $\mathcal{G}_{N(s)}$ is
contained in $\mathcal{G}_{i-1}$ as $(i-1) \geq N(s)$, along with the
\emph{tower property} and the fact that we can \emph{pull out what is known}
from the conditional expectation \cite{schilling05}. Since $X(t)$ is a
martingale, we have $ \mathbb{E}[X(t_i)\,|\,\mathcal{F}_{t_{i-1}}]
= X(t_{i-1})$ which means that
\begin{equation}
\mathbb{E} [Y(t_i^-) (X(t_i) - X(t_{i-1})) \,|\, \mathcal{G}_{N(s)}] = 0.
\end{equation}
Consequently, each term in the random sum vanishes and $\mathbb{E}[I(t)\,|\,
\mathcal{F}_s] = I(s)$. Summing up, if $X(t)$ is a martingale with respect to
$\mathcal{F}_t$ and if the integrand is bounded and predictable, one has that
$I(t)$ is also a martingale with respect to $\mathcal{F}_t$.

\section{Simulation}\label{sec:simulation}

In the previous section we have explicitly defined and rigorously characterized
a martingale stochastic integral driven by an uncoupled CTRW and given in
Eq.~(\ref{ito_integral}), as well as a more general class of stochastic
integrals given by Eq.~(\ref{stoch_integral}). A useful property of these
equations is that they can be easily implemented by means of Monte Carlo
simulation, as will be shown here for the case $Y(t) = X(t)$. The theory of
Sec.~\ref{sec:theory} is the basis for the Monte Carlo solution of stochastic
differential equations driven by CTRWs and discussed above in
Sec.~\ref{sec:motivation}.

The marginal distributions of jumps and waiting times presented in
Sec.~\ref{sec:motivation} are apparently demanding, but they can be sampled
easily using one-line transformation formulas \cite{fulger08,devroye86,
devroye96}. A random number $\xi$ drawn from the symmetric L\'evy
$\alpha$-stable probability density, Eq.~(\ref{levy}), can be obtained from two
independent uniform random numbers $U, V \in (0,1)$ through a transformation
due to Chambers, Mallows and Stuck \cite{chambers76,mcculloch96},
\begin{equation}\label{chambers}
\xi = \gamma_x\left(\frac{-\log U \cos\Phi}{\cos((1-\alpha)\Phi)}
\right)^{1-\frac{1}{\alpha}} \frac{\sin(\alpha\Phi)}{\cos\Phi},
\end{equation}
where $\Phi = \pi(V-1/2)$. For $\alpha = 2$ Eq.~(\ref{chambers}) reduces to
$\xi = 2\gamma_x\sqrt{-\log U}\,\sin\Phi$, i.e.\ the Box-Muller method for
Gaussian deviates with standard deviation $\sigma = \sqrt{2}\gamma_x$. A random
number $\tau$ drawn from the one-parameter Mittag-Leffler probability density,
Eq.~(\ref{mldensity}), can similarly be obtained from two independent uniform
random numbers $U, V \in (0,1)$ through a transformation proposed by Kozubowski
and Rachev \cite{kozubowski99,germano08}:
\begin{equation}\label{kozubowski}
\tau = -\gamma_t\log U \left(\frac{\sin(\beta\pi)}{\tan(\beta\pi V)}
-\cos(\beta\pi)\right)^\frac{1}{\beta}.
\end{equation}
For $\beta = 1$ Eq.~(\ref{kozubowski}) reduces to the transformation formula
for the exponential distribution, $\tau = -\gamma_t\log U$.

Now, as outlined above, the Monte Carlo simulation of an uncoupled CTRW is
straightforward. To compute the value $X(t)$, generate a sequence of $N(t)+1$
iid waiting times $\tau_i$ until their sum is greater than $t$. Discard the
last waiting time and generate $N(t)$ iid jumps $\xi_i$. Their sum is the
desired value of $X(t)$. Based on Eqs.~(\ref{ctrw}) and (\ref{renewal}), this
algorithm was used to generate Fig.~\ref{fig:walks}. This procedure is also the
basis to compute $I(t)$ according to Eq.~(\ref{ito_integral}), or more in
general $J_\vartheta(t)$ according to Eq.~(\ref{stoch_integral}), and the
covariation $[X,Y](t)$ according to Eq.~(\ref{covariation}). Each jump $\xi_i$
is multiplied by $Y(t_i^-)$, $(1-\vartheta)Y(t_i^-) + \vartheta Y(t_i)$, or
$Y(t_i)-Y(t_i^-)$, and the results of these multiplications are summed to
obtain respectively $I(t)$, $J_\vartheta(t)$ and $[X,Y](t)$. C++ code for the
case $Y(t) = X(t)$ can be found in the appendix.

Figs.~\ref{fig:dens1} and \ref{fig:dens2} show histograms from 1 million Monte
Carlo realizations of $X(t)$, $I(t) = \int_0^t X(s) \, dX(s)$, $S(t) = \int_0^t
X(s) \circ dX(s)$ and $[X](t)$, where $t = 1$ and $X(t)$ is a symmetric CTRW
with jump and time scale parameters linked by the relation $\gamma_x^\alpha
/\gamma_t^\beta = D = 1$. Thus the integrals in Figs.~\ref{fig:dens1} and
\ref{fig:dens2} give the Monte Carlo solution for $t = 1$ of the stochastic
differential equation $dZ = X dX$ with initial condition $Z(0) = 0$.
Since the It\=o integral is a martingale starting at zero, its mean is zero.
This is not true for the Stratonovich integral. The probability density of the
Stratonovich integral $S(t) = X^2(t)/2$ can be worked out from the density of
the stochastic process $X(t)$ by the transformation
$p_S(s,t) = \sum_i p_X(x_i(s),t)|dx_i(s)/ds|$,
where the sum is over all $x_i$ that yield the same $s$. For $s = x^2/2$ this
is $x_{1,2} = \pm \sqrt{2s}$ and thus
\begin{equation}\label{transformationXtoS}
p_S(s,t) = 2 p_X(\sqrt{2s},t) / \sqrt{2s}, \quad s > 0.
\end{equation}
In the diffusive limit the NCPP $X(t)$ approximates the Bachelier-Wiener
process $B(t)$ \cite{zygadlo03}, and thus the probability density of the
process $X(t)$ approximates the density of $B(t)$, Eq.~(\ref{wiener}).
The analytic probability density for the Stratonovich integral $S$ in the
diffusive limit can be obtained inserting the probability density of the
Bachelier-Wiener process into the transformation formula given by
Eq.~(\ref{transformationXtoS}), yielding
\begin{equation}
p_S(s,t;D) = \frac{1}{\sqrt{2 \pi Dts}}\exp\left(-\frac{s}{2Dt}\right),
\quad s > 0.
\end{equation}

\begin{figure*}[h]
\includegraphics[angle=-90,width=.95\columnwidth]{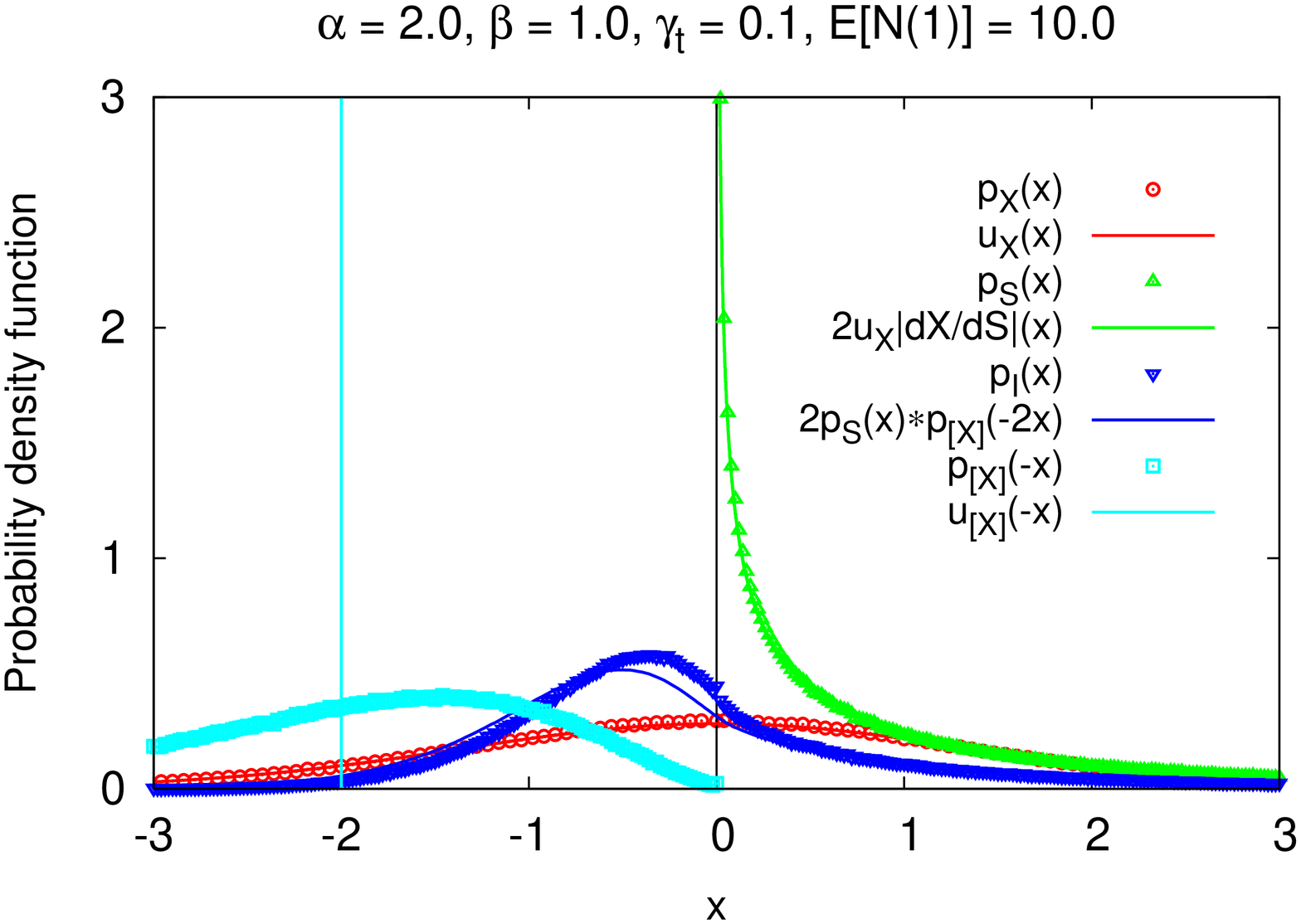}
\includegraphics[angle=-90,width=.95\columnwidth]{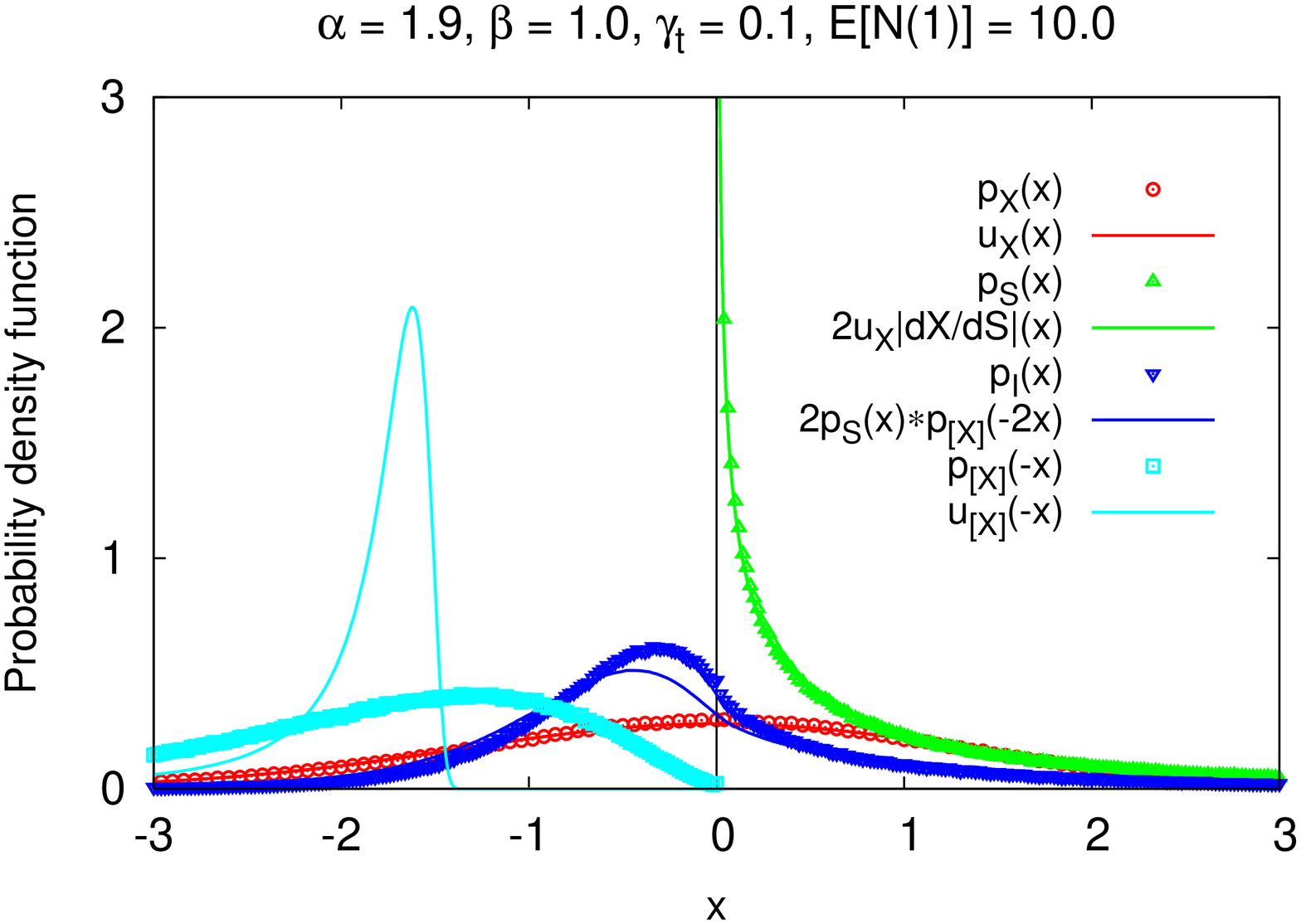}
\includegraphics[angle=-90,width=.95\columnwidth]{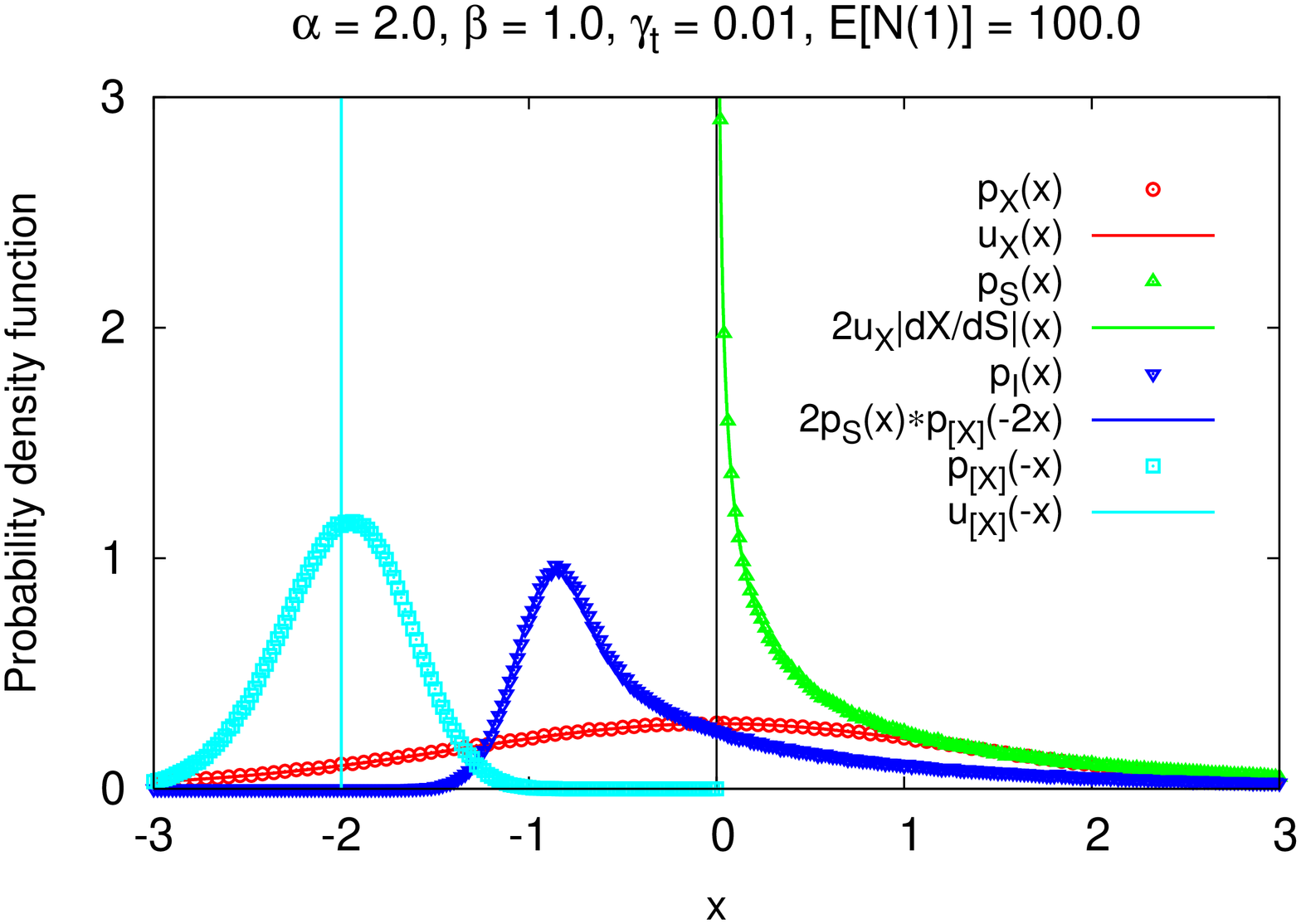}
\includegraphics[angle=-90,width=.95\columnwidth]{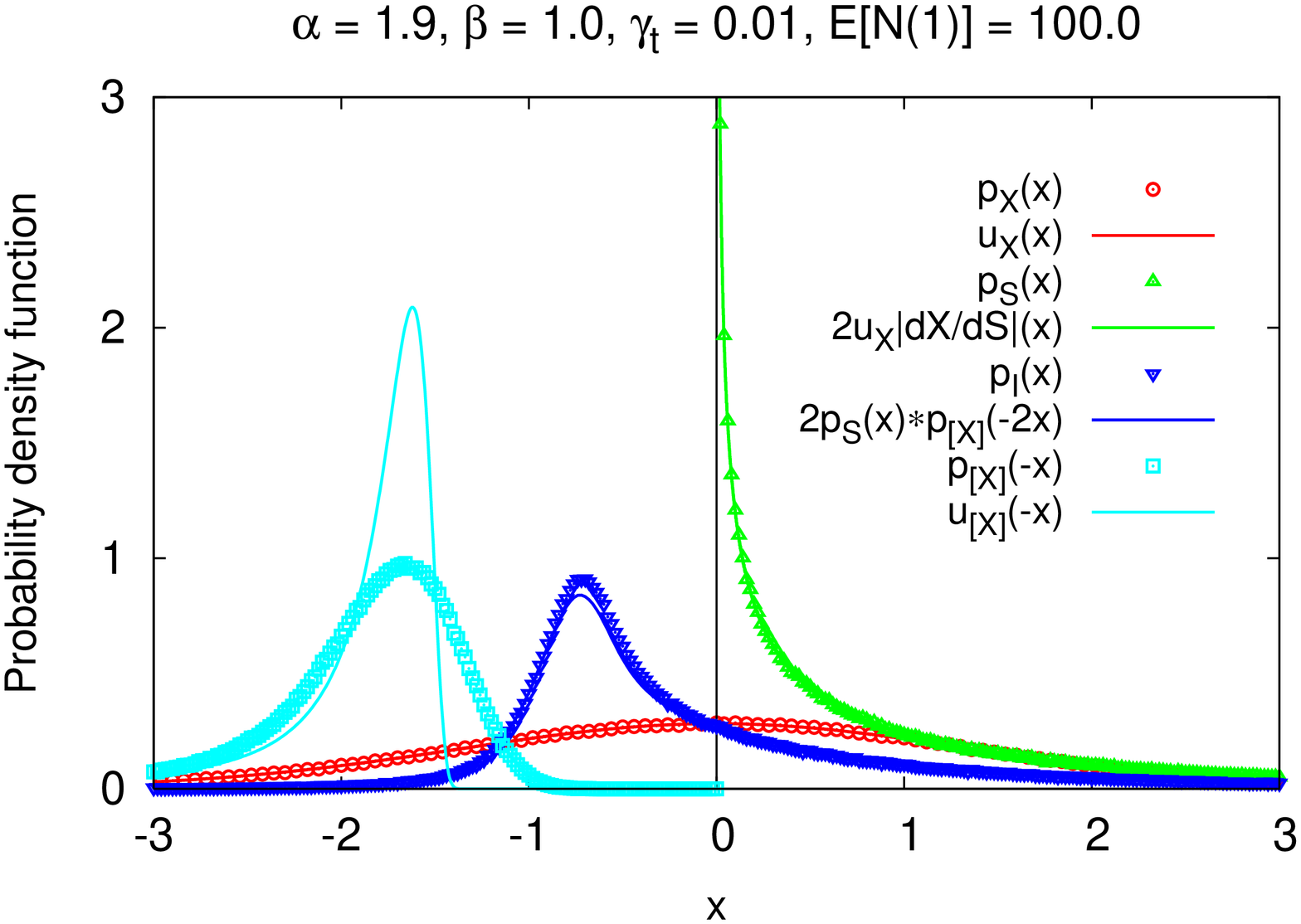}
\includegraphics[angle=-90,width=.95\columnwidth]{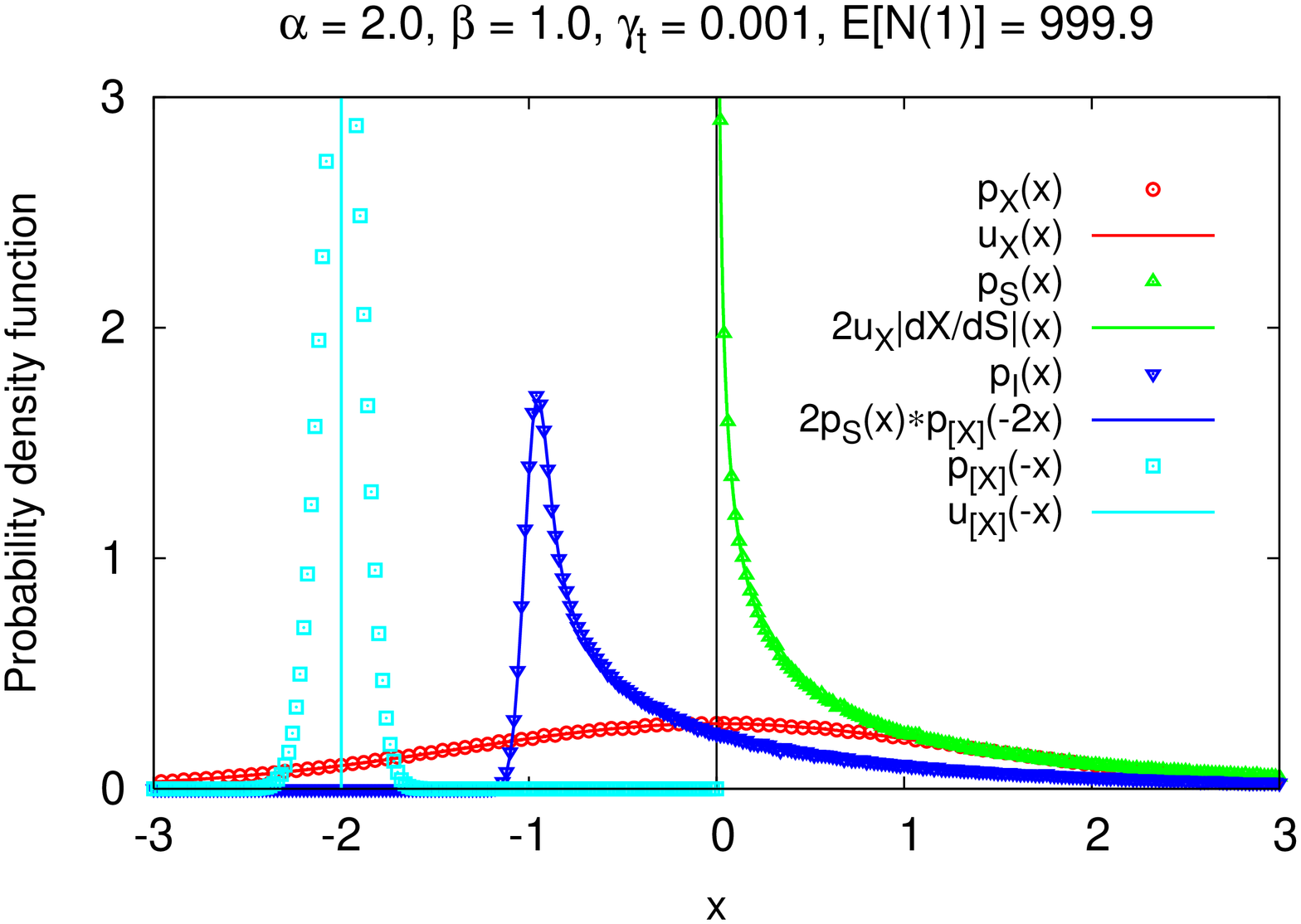}
\includegraphics[angle=-90,width=.95\columnwidth]{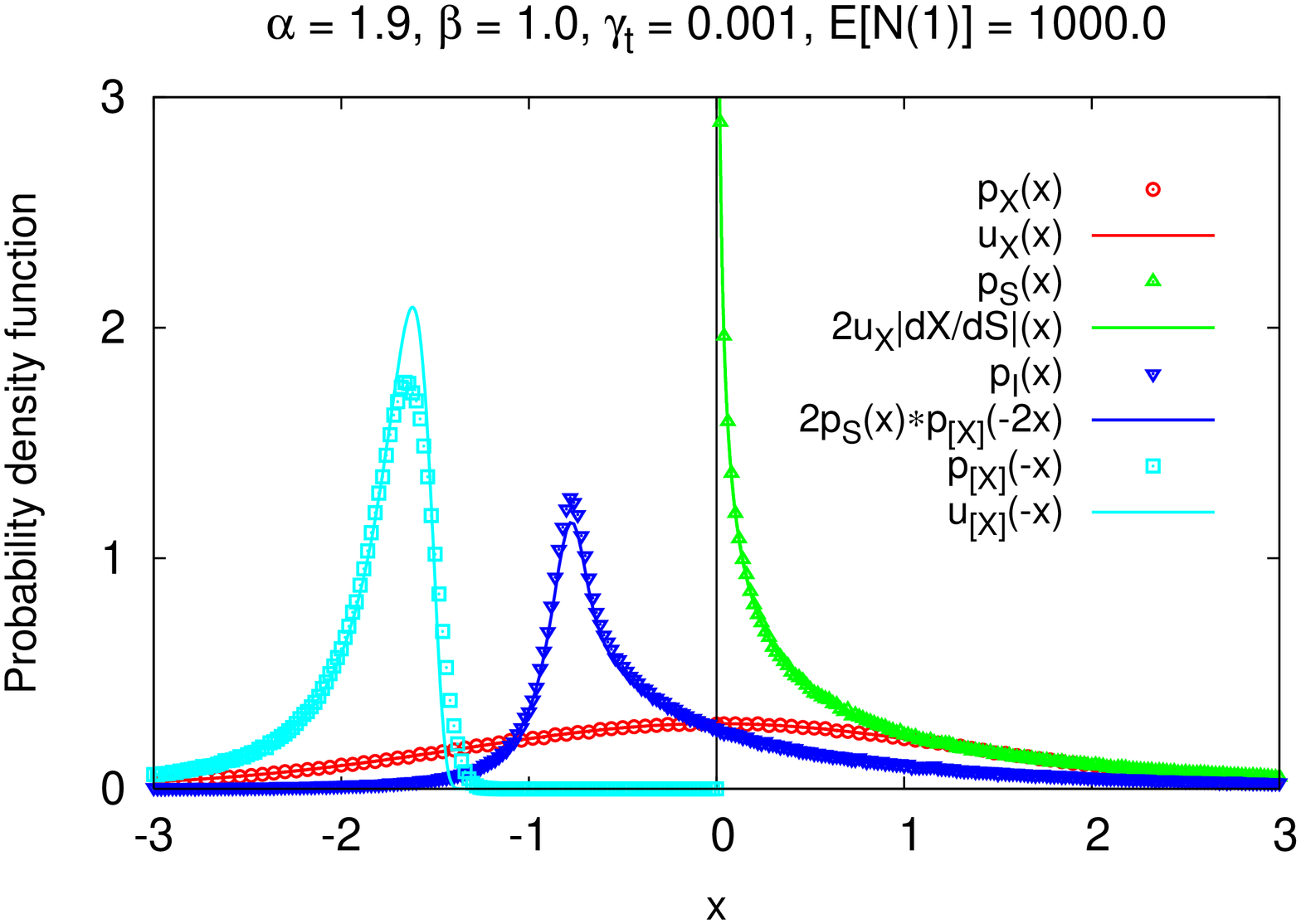}
\includegraphics[angle=-90,width=.95\columnwidth]{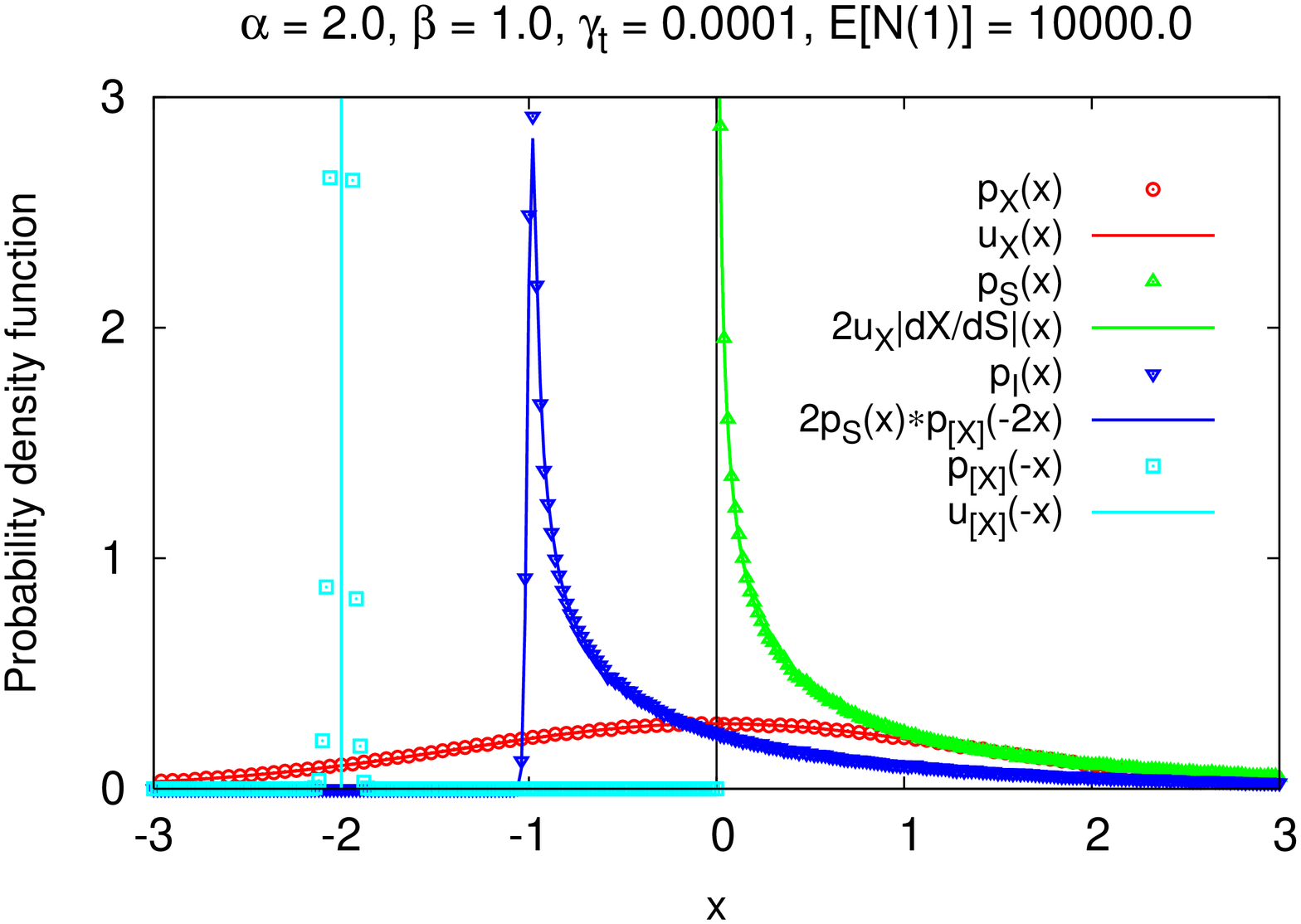}
\includegraphics[angle=-90,width=.95\columnwidth]{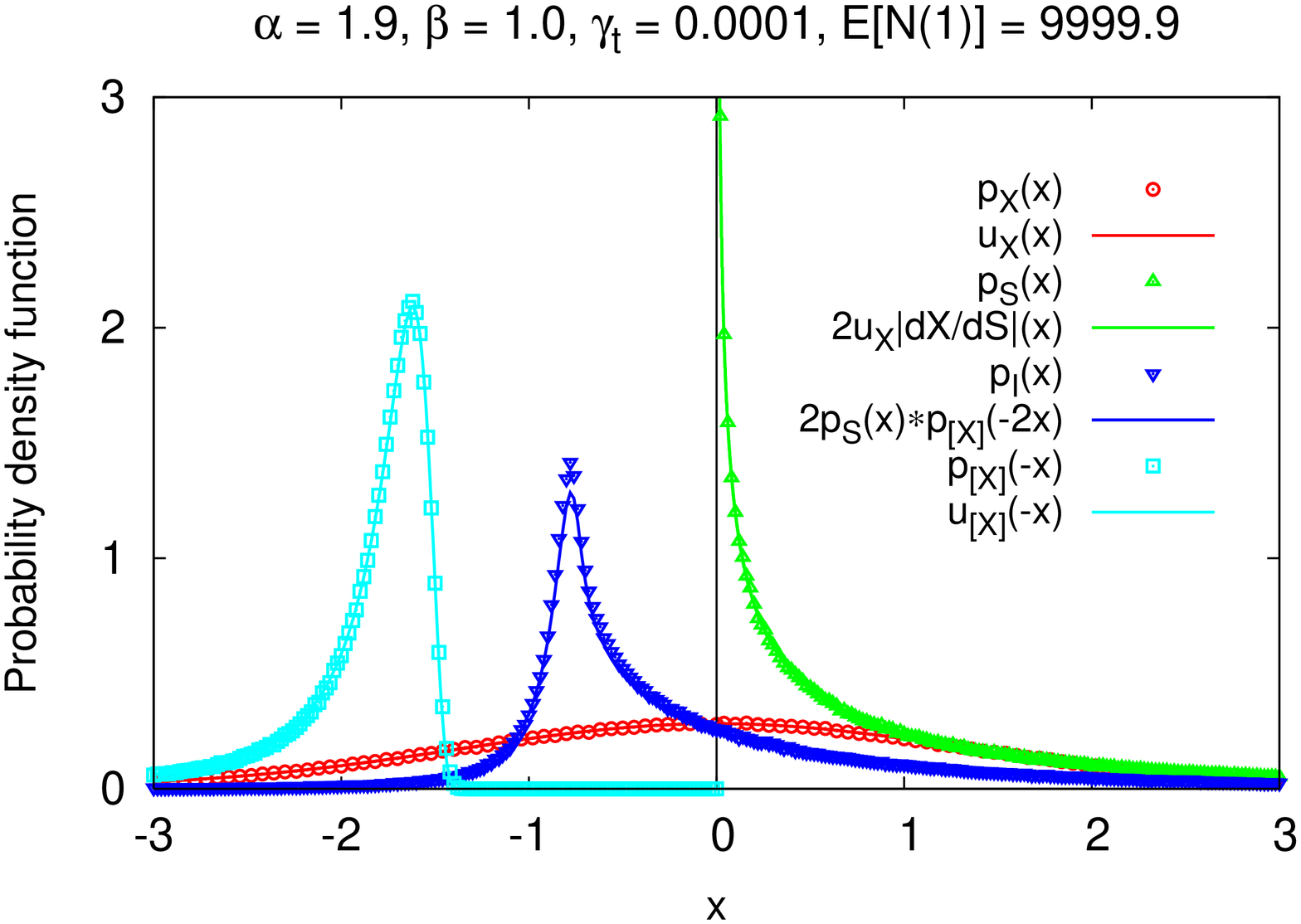}
\caption{\label{fig:dens1}
(Color online) Convergence of the empirical probability densities $p$ from
1~million Monte Carlo runs (points) to the analytic probability densities $u$
(lines) in the diffusive limit for a CTRW $X(t)$, its Stratonovich integral
$S(t)$, its It\=o integral $I(t)$, and its quadratic variation $[X](t)$,
with $t = 1$ and different choices of the index parameters $\alpha, \beta$
and of the scale parameters $\gamma_x, \gamma_t$, where
$\gamma_x^\alpha/\gamma_t^\beta = D = 1$. $\mathbb{E}[N(t)]$ is the average
number of jumps per run.}
\end{figure*}

\begin{figure*}[h]
\includegraphics[angle=-90,width=.95\columnwidth]{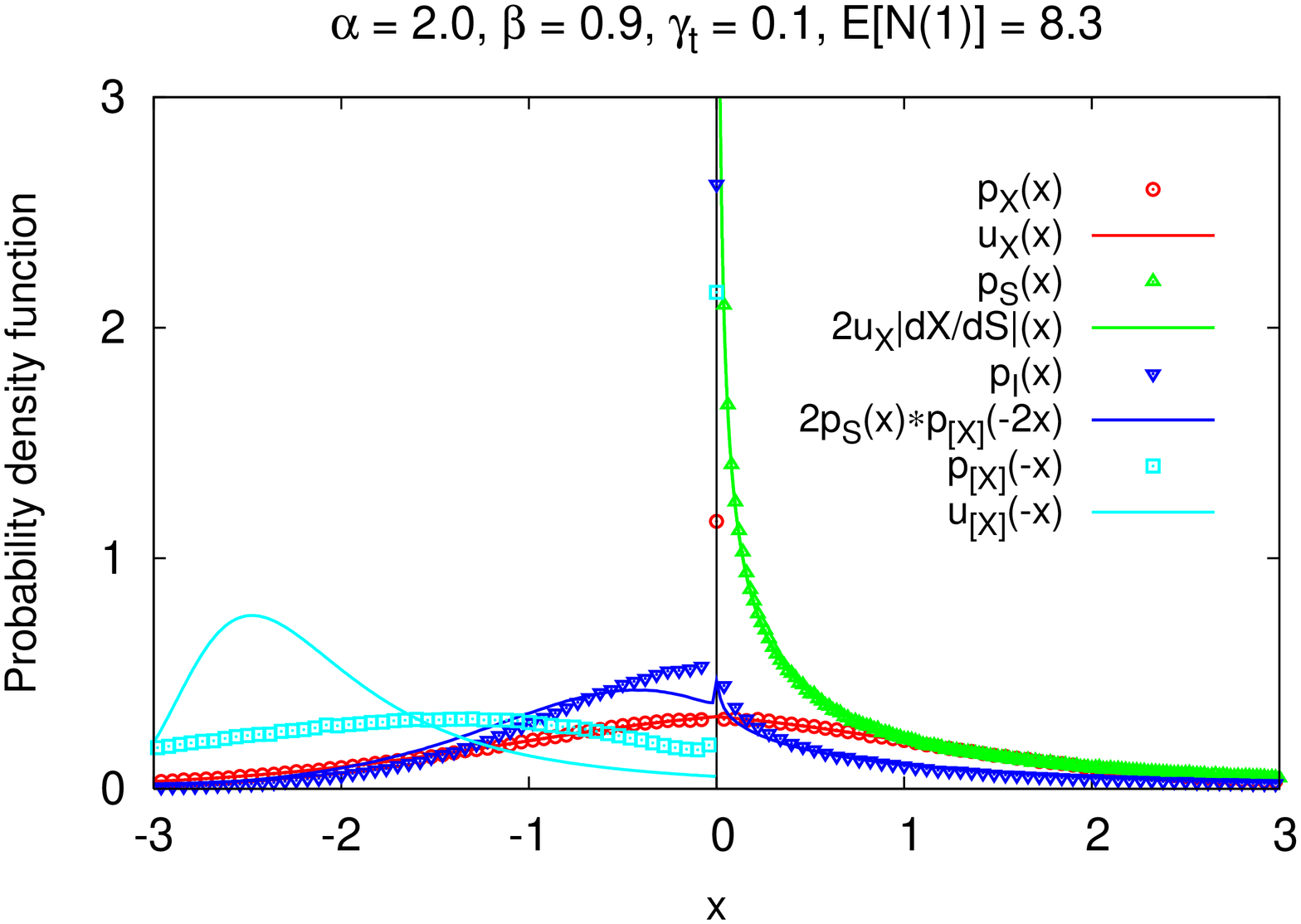}
\includegraphics[angle=-90,width=.95\columnwidth]{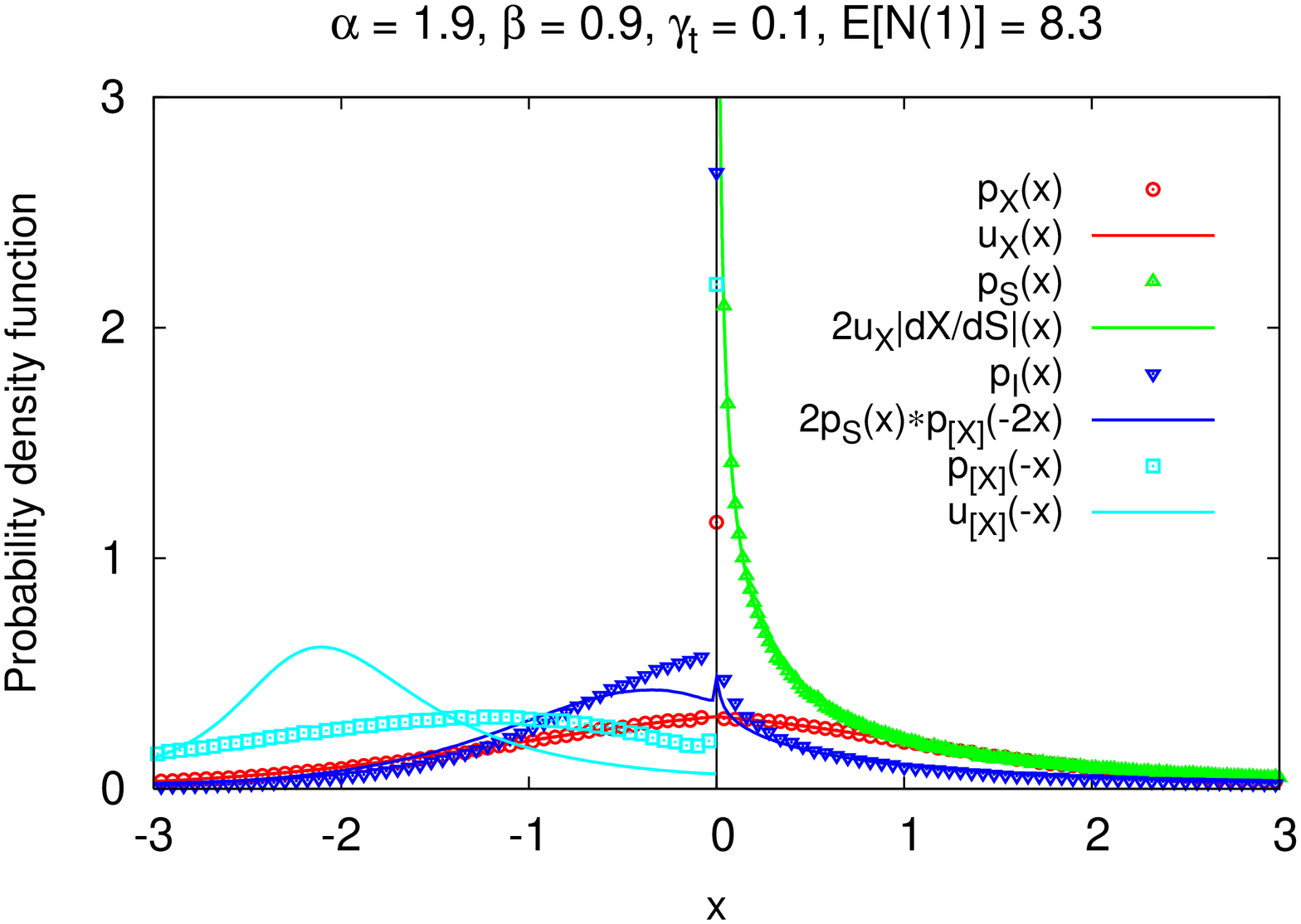}
\includegraphics[angle=-90,width=.95\columnwidth]{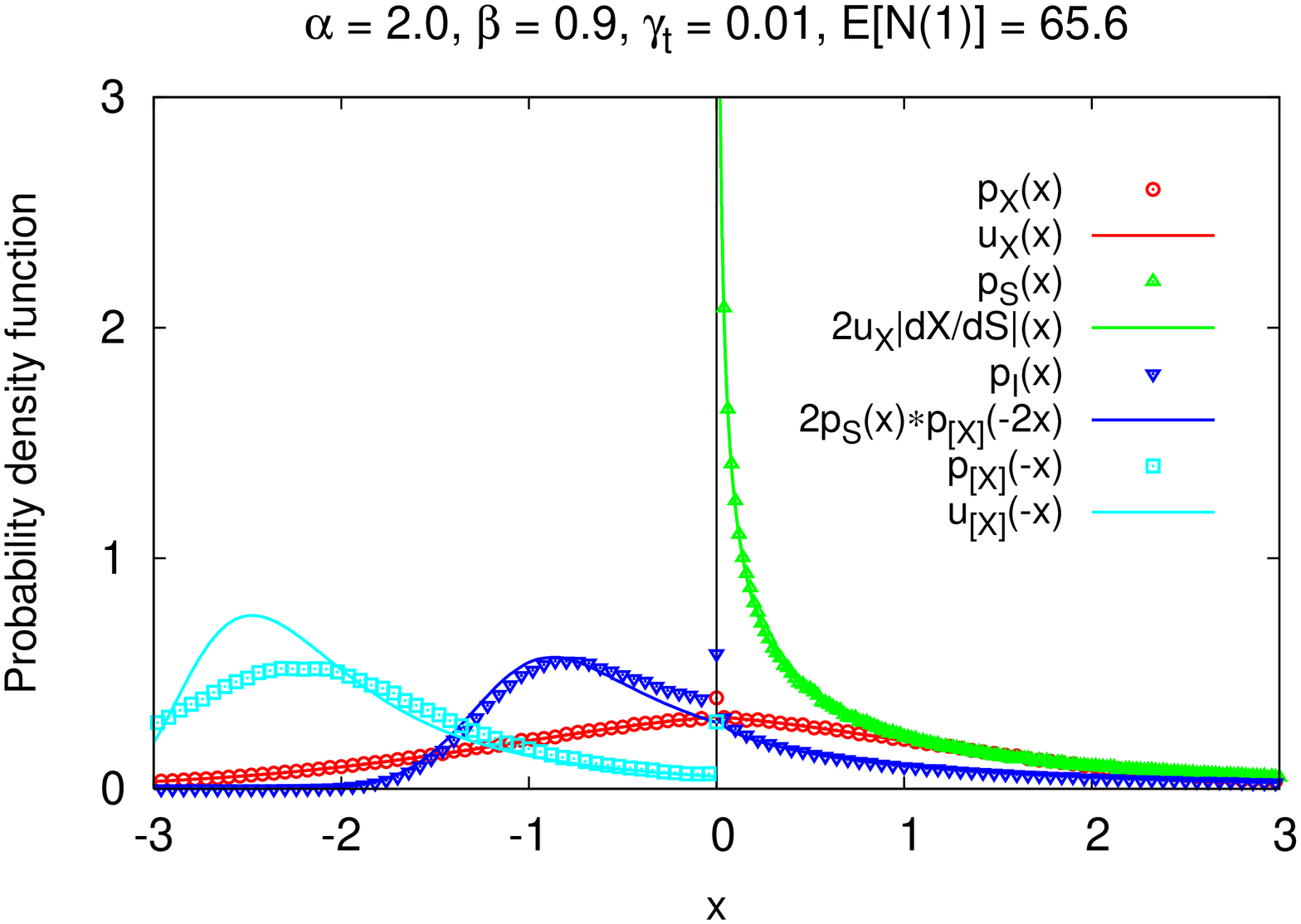}
\includegraphics[angle=-90,width=.95\columnwidth]{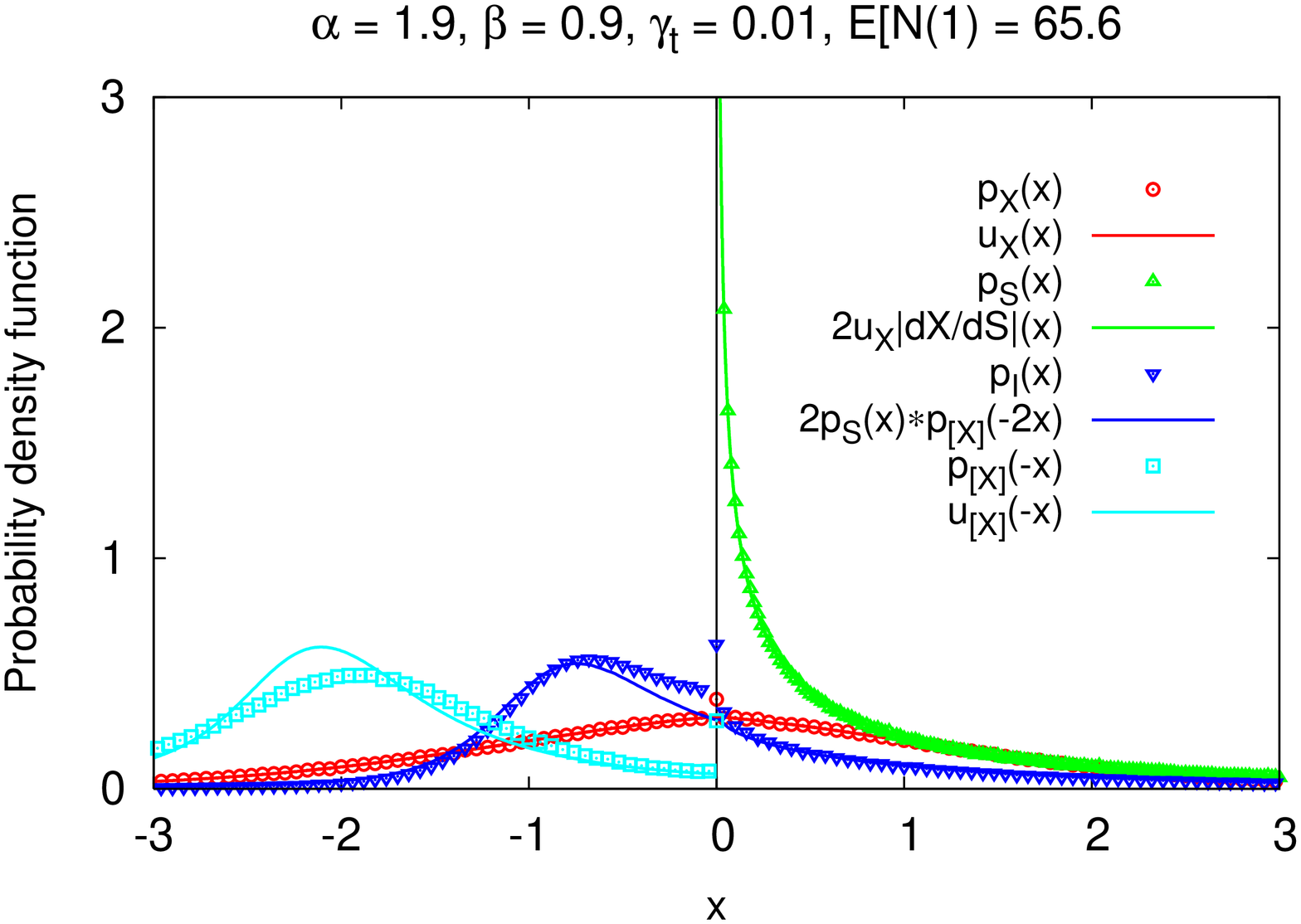}
\includegraphics[angle=-90,width=.95\columnwidth]{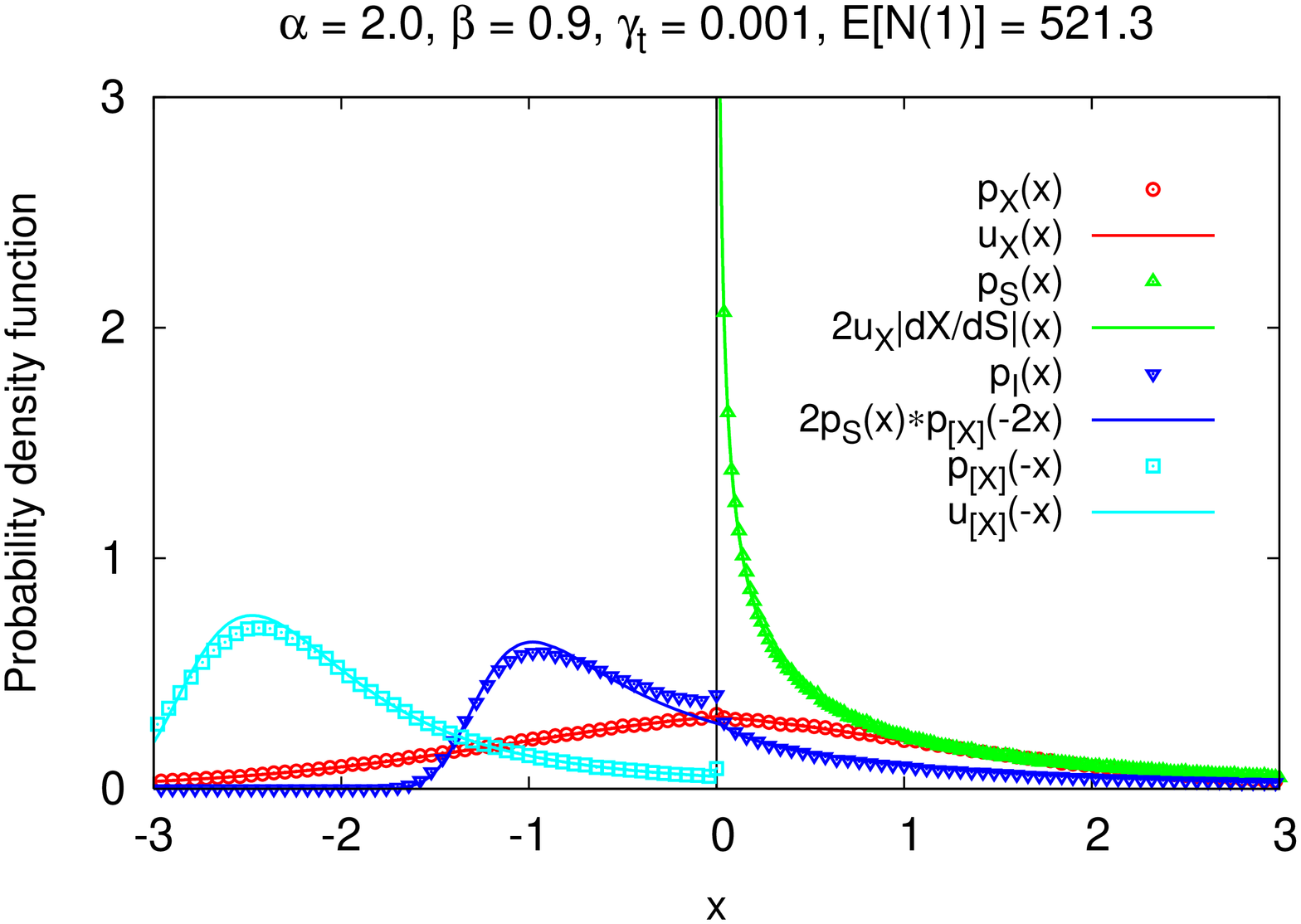}
\includegraphics[angle=-90,width=.95\columnwidth]{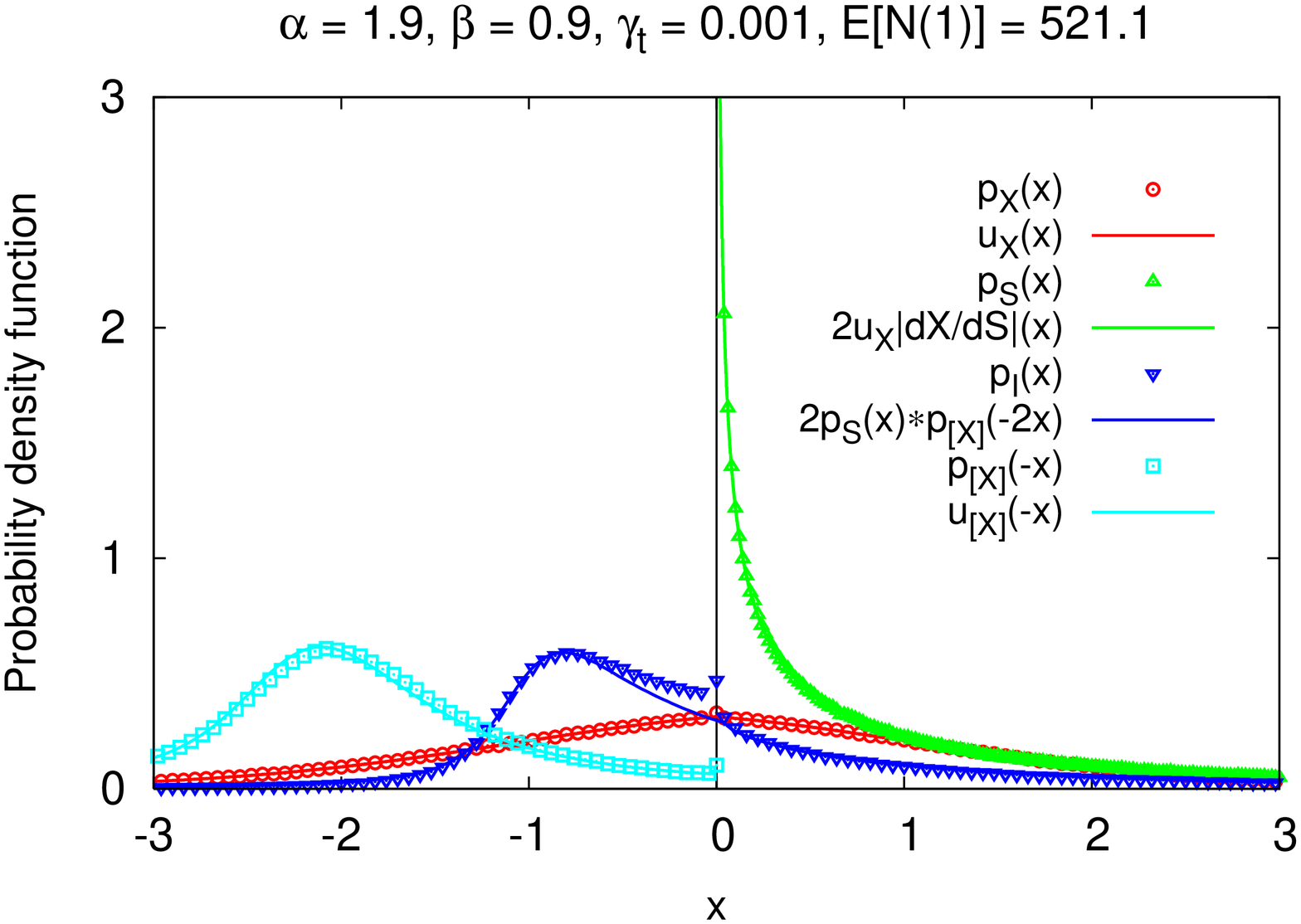}
\includegraphics[angle=-90,width=.95\columnwidth]{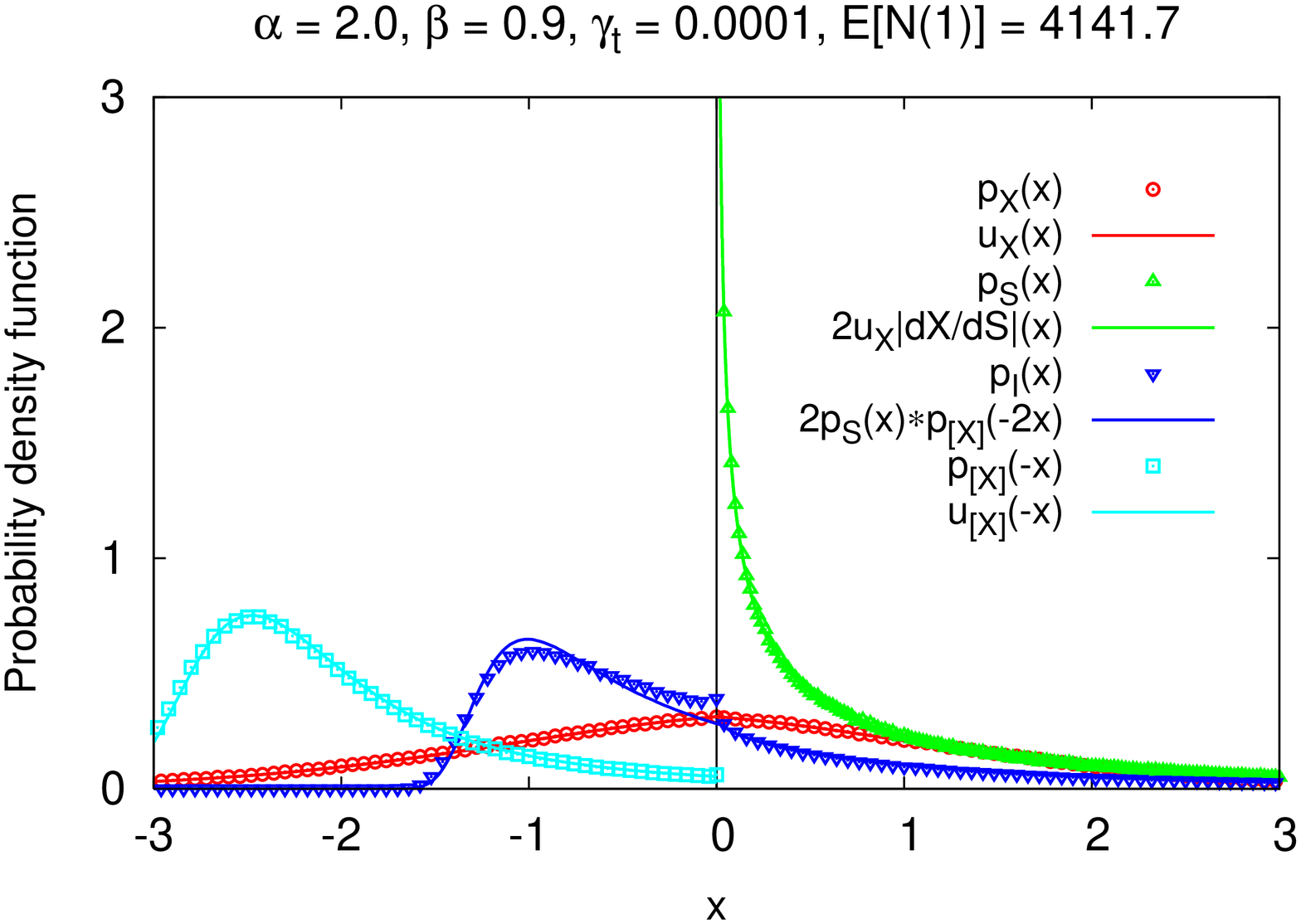}
\includegraphics[angle=-90,width=.95\columnwidth]{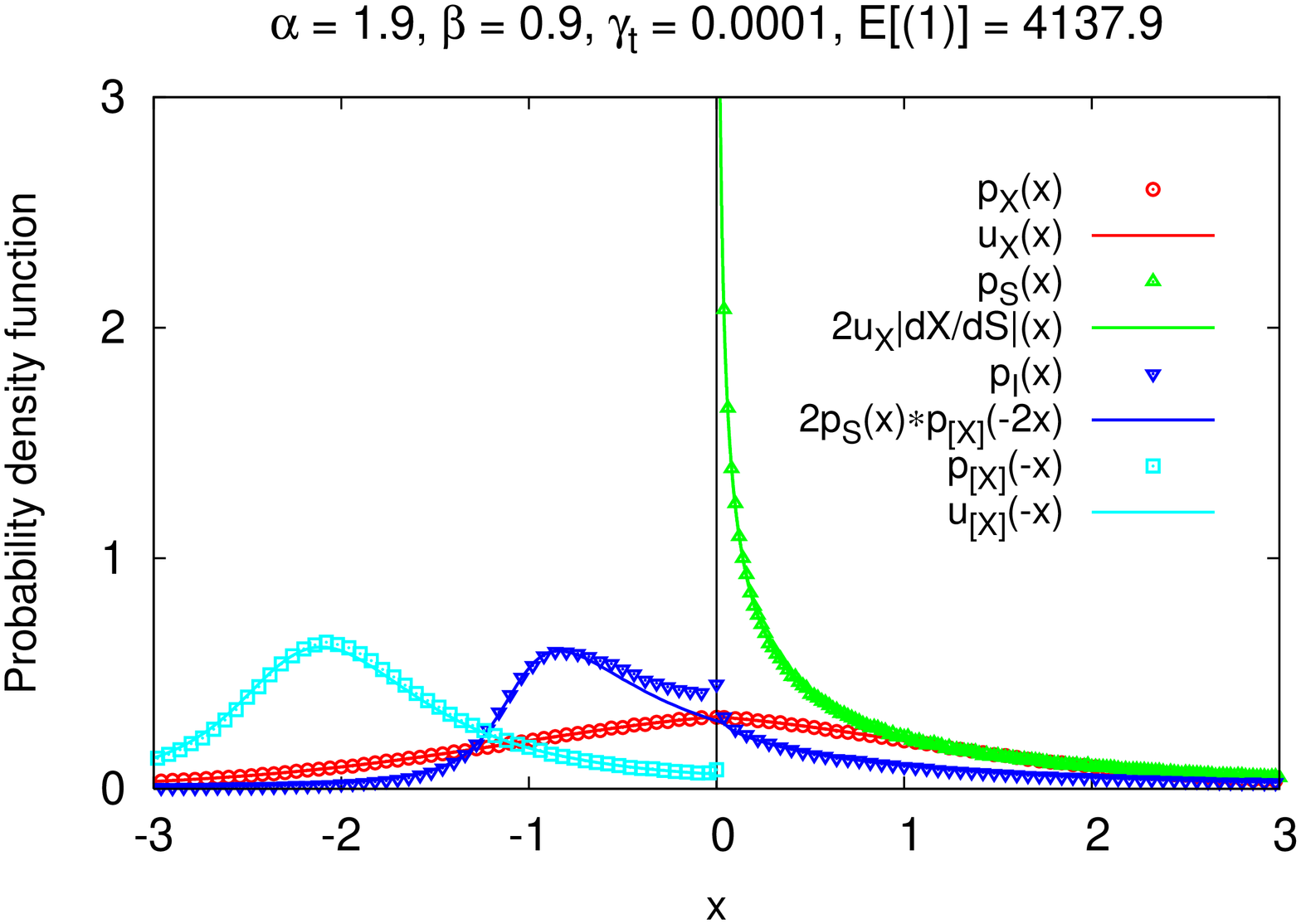}
\caption{\label{fig:dens2}
(Color online) Convergence of the empirical probability densities $p$ from
1~million Monte Carlo runs (points) to the analytic probability densities $u$
(lines) in the diffusive limit for a CTRW $X(t)$, its Stratonovich integral
$S(t)$, its It\=o integral $I(t)$, and its quadratic variation $[X](t)$,
with $t = 1$ and different choices of the index parameters $\alpha, \beta$
and of the scale parameters $\gamma_x, \gamma_t$, where
$\gamma_x^\alpha/\gamma_t^\beta = D = 1$. $\mathbb{E}[N(t)]$ is the average
number of jumps per run.}
\end{figure*}

According to Eq.~(\ref{compensator}) here $I(t) = S(t) - [X](t)/2$; if the
dependence of $S$ and $[X]$ is small, the probability density of the It\=o
integral is approximated by the convolution of the probability density of the
Stratonovich integral with that of the quadratic variation mirrored around zero
and scaled to half its width:
\begin{equation}\label{convolution}
p_I(x,t) \simeq 2 \int_{-\infty}^{+\infty} p_S(x+2x',t) p_{[X]}(-2x',t) \,dx'.
\end{equation}

For all choices of $\alpha$ and $\beta$ the agreement between the analytic
expressions for $X(t)$ and $S(t)$ in the diffusive limit and the empirical
results from Monte Carlo simulation of the CTRWs is fair already for the
largest value $\gamma_t = 0.1$: the curves cannot be distinguished by eye at
the scale of our plots. Therefore we did not evaluate the analytic probability
density for $X(t)$, Eq.~(\ref{NCPPdens}), available for the particular case of
a NCPP only, i.e.\ the left column of Fig.~\ref{fig:dens1}.
Instead the quadratic variation $[X](t)$ and consequently the It\=o integral
tend visibly more slowly to their diffusive limits. For a NCPP the diffusive
limit of $[X](t)$ is $[B](t) = 2Dt$. In this limit $I(t) = S(t) - Dt = B^2(t)/2
- Dt$, corresponding to the well-known result that the probability density of
the It\=o integral is equal to the density of the Stratonovich integral shifted
by $-Dt$, i.e.\ $p_I(x,t) = p_S(x+Dt,t)$. Though the quadratic variation of the
NCPP is appreciably different from its limit $\delta(x-2Dt)$, where $Dt = 1$,
for any non infinitesimal value of $\gamma_t$ as shown in the left column of
Fig.~\ref{fig:dens1}, for $\gamma_t = 0.01$ there is a good
agreement between the It\=o integrals from Monte Carlo and from
Eq.~(\ref{convolution}).

The density of the quadratic variation for a CTRW can be obtained 
from the density of squared jumps, $\lambda_{\xi^2}(x)$, that results from
a transformation of the density of jumps, $\lambda_\xi(\sqrt{x})$, similar to
the one that leads from $p_X(x,t)$ to $p_S(x,t)$,
Eq.~(\ref{transformationXtoS}), except for a factor 2:
\begin{equation}\label{transformationXtoQ}
\lambda_{\xi^2}(x;\gamma_x) = \lambda_\xi(\sqrt{x};\gamma_x)/\sqrt{x}
\end{equation}
Inserting this equation into the solution of the Montroll-Weiss equation in the
space-time domain, Eq.~(\ref{MWsolution}), gives
\begin{equation}\label{CTRWqvdens}
p_{[X]}(x,t;\gamma_x,\gamma_t) = \sum_{n=0}^\infty p_N(n,t;\gamma_t)
\lambda_{\xi^2}^{*n}(x;\gamma_x),
\end{equation}
where $x > 0$. Unfortunately even for an NCPP the $n$-fold convolution cannot
be computed as easily as for $p_X(x,t)$ in Eq.~(\ref{NCPPdens}).
However, the characteristic function of the quadratic variation can be written
as 
\begin{equation}\label{CTRWqvcharfun}
\widehat{p}_{[X]}(k,t;\gamma_x,\gamma_t)
= \sum_{n=0}^\infty p_N(n,t;\gamma_t) \widehat{\lambda}_{\xi^2}^n(k;\gamma_x).
\end{equation}
In order to consider non-exponential waiting times with power-law tails and
infinite first moment, for the sake of simplicity let us assume that $p_N(n,t)$
is the distribution of the Mittag-Leffler counting process \cite{scalas04},
\begin{equation}
p_N(n,t;\gamma_t) = \frac{(t/\gamma_t)^{\beta n}}{n!}
E_{\beta}^{(n)}\left(-(t/\gamma_t)^\beta\right),
\end{equation}
where
\begin{equation}
E_{\beta}^{(n)}(z) = \frac{d^n}{dz^n} E_{\beta} (z).
\end{equation}
This choice is more general than it seems, as the Mittag-Leffler distribution
for waiting times is an attractor for the thinning procedure used to obtain the
diffusive limit \cite{mainardi04}. Using the Mittag-Leffler distribution from
the beginning simplifies the derivation of this limit.
Then Eq.~(\ref{CTRWqvcharfun}) becomes \cite{scalas06}
\begin{equation}\label{CTRWqvcharfun2}
\widehat{\lambda}_{[X]}(k,t;\gamma_x,\gamma_t) = E_\beta \big(
-(t/\gamma_t)^\beta (1-\widehat{\lambda}_{\xi^2}(k;\gamma_x)) \big).
\end{equation}
As the jumps $\xi$ follow a L\'evy $\alpha$-stable distribution, for
$x \rightarrow \infty,\ \lambda_{\xi^2}(x;\gamma_x) \sim
(x/\gamma_x)^{-\alpha/2-1}$, and the sum of $\xi_i^2$ converges to the positive
stable distribution with index $\alpha/2$, whose characteristic function is
\begin{equation}
\widehat{\lambda}_{\xi^2}(k;\gamma_x) = \widehat{L}_{\alpha/2}^+(k;\gamma_x)
\equiv \exp \big( (-i\gamma_x^2 k)^{\alpha/2} \big).
\end{equation}
The scale parameter $\gamma_x$ is the same as in the L\'evy stable
distribution, Eq.~(\ref{levy}). 
Inserting this distribution in Eq.~(\ref{CTRWqvcharfun2}), the diffusive limit
yields the following characteristic function for the quadratic variation:
\begin{equation}\label{CTRWqvcharfun3}
\widehat{u}_{[X]}(k,t;D) = E_\beta\big(-D(-ik)^{\alpha/2}t^\beta\big).
\end{equation}
Now we can proceed in a similar fashion as for the solution of the FDE,
Eqs.~(\ref{scalingfunction}--\ref{greenfunction}). Defining $\kappa =
kt^{2\beta/\alpha}$ and 
\begin{equation}\label{scalaswright}
M_{\alpha,\beta}(\xi;D) = \mathcal{F}^{-1}_\kappa
\big[E_\beta\big(-D(-i\kappa)^{\alpha/2}\big)\big](\xi),
\end{equation}
where $\quad \xi > 0$, we obtain the quadratic variation for the diffusive
limit in the space-time domain,
\begin{equation}\label{qvdens}
u_{[X]}(x,t;D) = t^{-2\beta/\alpha}\,M_{\alpha,\beta}(xt^{-2\beta/\alpha};D).
\end{equation}
When $\alpha = 2,\ M_{2,\beta}(\xi)$ coincides with the right half of the
Mainardi-Wright function \cite{mainardi01}, which is also called M-function of
Wright type because its shape recalls a capital M centered in the origin.
When $\alpha = 2$ and $\beta = 1$ (standard diffusion case), a delta function
$u_{[X]}(x,t;D) = \delta(x-2Dt)$ is recovered, corresponding to the quadratic
variation of the Bachelier-Wiener process, $[X](t) = 2 D t$.
The plots in Figs.~\ref{fig:dens1} and \ref{fig:dens2} display quadratic
variations both from Monte Carlo and from Eq.~(\ref{qvdens}). The convergence
of the quadratic variation in the diffusive limit can be used to prove that the
integrals of $X(t)$ as defined in Sec.~\ref{sec:theory} converge.




\section{Conclusions and outlook}

This paper is based on the definition, given in Eq.~(\ref{stoch_integral}),
of a class of stochastic integrals $J_\vartheta(t)$ driven by a CTRW $X(t)$.
For $\vartheta = 0$ this results in the It\=o integral $I(t)$,
Eq.~(\ref{ito_integral}), for $\vartheta = 1/2$ in the Stratonovich integral.
If the process $X(t)$ that defines the measure used in Eq.~(\ref{ito_integral})
is a martingale with respect to its natural filtration, then $I(t)$ is a
martingale too; this is a consequence of the martingale transform theorem.
It turns out that an uncoupled CTRW with zero-mean jumps is a martingale.
The stochastic integration theory developed here is more general than the one
sketched in Ref.~\cite{zygadlo03}, as it can be applied also to a CTRW that is
neither Markovian nor L\'evy. In fact, exponential waiting times are not needed
to prove that $I(t)$ is a martingale if $X(t)$ is a martingale.

The theory presented in Sec.~\ref{sec:theory} lies at the foundation of the
Monte Carlo method for integrating stochastic differential equations driven by
CTRWs. As explained in Sec.~\ref{sec:introduction}, these results are relevant
for applications in physics and economics as well as in all those fields like
insurance and finance where martingale methods can help in the quantitative
evaluation of risk. Eq.~(\ref{stoch_integral}) is a convenient basis for the
Monte Carlo calculation of stochastic integrals. This is shown in
Sec.~\ref{sec:simulation}, where Monte Carlo realizations of CTRWs are used to
effectively approximate the It\=o and Stratonovich integrals driven by the
Bachelier-Wiener process and, more generally, by the solution of the space-time
fractional diffusion equation.

We believe that up-to-date mathematical methods from probability theory and
stochastic calculus are beneficial to the study of the CTRW and of other random
processes useful in statistical physics. We fear that progress will be slower
or impossible if these methods are ignored by physicists.

Future work will deal with Monte Carlo simulations for coupled CTRWs where
jumps and waiting times obey fat-tailed distributions \cite{meerschaert02,
meerschaert06}. There will also be a discussion of convergence based on the
results collected in \cite{jacod03}.

\section*{Acknowledgements}

E.~S.\ had inspiring discussions with F.~Mainardi, who pointed him to Wright
functions, R.~Gorenflo and F.~Rapallo. His visits in Marburg were funded
through a grant by East Piedmont University. The stay of M.~P.\ in Marburg was
supported by two DAAD grants. G.~G.\ benefitted from listening to lectures on
stochastic integration by D.~Sondermann during the first year of his graduate
studies.

\section*{Appendix}

Below are salient lines from the central loop of our C++ program for the
Monte Carlo calculation of a CTRW $X(t)$, its quadratic variation $[X(t)]$, its
It\=o integral $I(t)$ and its Stratonovich integral $S(t)$ as described in
Sec.~\ref{sec:simulation} and shown in Figs.~\ref{fig:dens1}--\ref{fig:dens2}.

\begin{verbatim}
jumps = 0

// Loop over runs
for (run = 1; run <= runs; run++) {

    // Initialize and increment t, x, etc.
    t = 0, x = 0, qvar = 0, ito = 0, str = 0,
    tau = random.t();          // Eq. (47)
    while (t + tau < t_max) {
        t += tau;              // time t
        xi = random.x();       // Eq. (46)
        qvar += xi*xi;         // [X(t)]
        ito += x*xi;           // I(t)
        str += (x+xi/2)*xi;    // S(t)
        x += xi;               // X(t)
        tau = random.t();      // Eq. (47)
        jumps++;               // N(t)
    }

    // Update histograms at the end of each run
    hisx.add(x);               // X(t)
    hisq.add(qvar);            // [X(t)]
    hisi.add(ito);             // I(t)
    hiss.add(str);             // S(t)
}
\end{verbatim}

CPU times grow linearly with the number of jumps $N(t)$ and take 1--3 $\mu$sec
per jump depending on $\alpha$ and $\beta$ on a 2.2 GHz AMD Athlon 64 X2
``Toledo'' Dual-Core processor with Fedora Core 7 Linux, using the \texttt{Ran}
uniform random number generator \cite{press07} and the GNU C++ compiler (g++)
version 4.1.2 with the -O3 -static optimization options.


\begin{thebibliography}{99}

\bibitem{montroll65}
E. Montroll and G. H. Weiss, J. Math. Phys. \textbf{6}, 167 (1965).

\bibitem{scher73a}
H. Scher and M. Lax, Phys. Rev. B \textbf{7}, 4491 (1973).

\bibitem{scher73b}
H. Scher and M. Lax, Phys. Rev. B \textbf{7}, 4502 (1973).

\bibitem{montroll73}
E. W. Montroll and H. Scher, J. Stat. Phys. \textbf{9}, 101 (1973).

\bibitem{scher75}
H. Scher and E. Montroll, Phys. Rev. B \textbf{12}, 2455 (1975).

\bibitem{shlesinger96}
M. F. Shlesinger, \textit{Random processes}, in \textit{Encyclopedia of Applied
Physics}, Vol.\ 16, edited by G. L. Trigg (VCH Publishers, New York, 1996),
pp.~45--70.

\bibitem{weiss94}
G. H. Weiss, \textit{Aspects and Applications of the Random Walk}
(North-Holland, Amsterdam, 1994).

\bibitem{metzler00}
R. Metzler and J. Klafter, Phys. Rep. \textbf{339}, 1 (2000).

\bibitem{metzler04}
R. Metzler and J. Klafter, J. Phys. A: Math. Gen. \textbf{37}, R161 (2004).

\bibitem{fulger08}
D. Fulger, E. Scalas, and G. Germano, Phys. Rev. E \textbf{77}, 021122 (2008).

\bibitem{meerschaert01}
M. M. Meerschaert and H. P. Scheffler, \textit{Limit Distributions for Sums
of Independent Random Vectors: Heavy Tails in Theory and Practice}
(Wiley, New York, 2001).

\bibitem{cox67}
D. R. Cox, \textit{Renewal Theory} (Methuen, London, 1967). 

\bibitem{feller71}
W. Feller, \textit{An Introduction to Probability Theory and its Applications}, Vol. 2 (Wiley, New York, 1971).

\bibitem{cox79}
D. R. Cox and V. Isham, \textit{Point Processes} (Chapman \& Hall, London,
1979).

\bibitem{billingsley79}
P. Billingsley, \textit{Probability and Measure} (Wiley, New York, 1979).

\bibitem{hoel72}
P. G. Hoel, S. C. Port, and J. Stone, \textit{Introduction to Stochastic
Processes} (Houghton Mifflin, Boston, 1972).

\bibitem{cinlar75}
E. \c{C}inlar, \textit{Introduction to Stochastic Processes} (Prentice-Hall,
Englewood Cliffs, 1975).

\bibitem{flomenbom05}
O. Flomenbom, J. Klafter, Phys. Rev. Lett. \textbf{95}, 098105 (2005).

\bibitem{flomenbom07}
O. Flomenbom, R. J. Silbey, Phys. Rev. E \textbf{76}, 041101 (2007).

\bibitem{janssen07}
J. Janssen and R. Manca, \textit{Semi-Markov Risk Models for Finance, Insurance
and Reliability} (Springer, New York, 2007).

\bibitem{bertoin96}
J. Bertoin, \textit{L\'evy Processes} (Cambridge University Press, Cambridge,
UK, 1996).

\bibitem{sato99}
K.-I. Sato, \textit{L\'evy Processes and Infinitely Divisible Distributions}
(Cambridge University Press, Cambridge, UK, 1999).

\bibitem{shlesinger74}
M. F. Shlesinger, J. Stat. Phys. \textbf{10}, 421 (1974).

\bibitem{tunaley74}
J. K. E. Tunaley, J. Stat. Phys. \textbf{11}, 397 (1974).

\bibitem{tunaley75}
J. K. E. Tunaley, J. Stat. Phys. \textbf{12}, 1 (1975).

\bibitem{tunaley76}
J. K. E. Tunaley, J. Stat. Phys. \textbf{14}, 461 (1976).

\bibitem{shlesinger82}
M. F. Shlesinger, J. Klafter, and Y. M. Wong, J. Stat. Phys. \textbf{27},
499 (1982).

\bibitem{benavraham00}
D. ben-Avraham and S. Havlin, \textit{Diffusion and Reactions in Fractals and
Disordered Systems} (Cambridge University Press, Cambridge, UK, 2000).

\bibitem{balakrishnan85}
V. Balakrishnan, Physica A \textbf{132}, 569 (1985).

\bibitem{hilfer95}
R. Hilfer and L. Anton, Phys. Rev. E \textbf{51}, R848 (1995).

\bibitem{scalas04}
E. Scalas, R. Gorenflo, and F. Mainardi, Phys. Rev. E \textbf{69}, 011107
(2004).

\bibitem{scalas06}
E. Scalas, Physica A \textbf{362}, 225 (2006).

\bibitem{negrete05}
D. del-Castillo-Negrete, B. A. Carreras, and V. E. Lynch, Phys. Rev. Lett.
\textbf{94}, 065003 (2005).

\bibitem{dubbeldam07a}
J. L. A. Dubbeldam, A. Milchev, V. G. Rostiashvili, and T. A. Vilgis,
Phys. Rev. E \textbf{76}, 010801 (2007).

\bibitem{dubbeldam07b}
J. L. A. Dubbeldam, A. Milchev, V. G. Rostiashvili, and T. A. Vilgis,
Europhys. Lett. \textbf{79}, 18002 (2007).

\bibitem{masoliver06}
J. Masoliver, M. Montero, J. Perell\'o, and G. H. Weiss, J. Econ. Behav.
Organ. \textbf{61}, 577 (2006).

\bibitem{embrechts97}
P. Embrechts, C. Kl\"uppelberg, and T. Mikosch, \textit{Modelling Extremal
Events for Insurance and Finance} (Springer, New York, 1997).

\bibitem{cartea07}
\'{A}. Cartea and D. del-Castillo-Negrete, Phys. Rev. E \textbf{76}, 041105
(2007).

\bibitem{zygadlo03}
R. Zygad{\l}o, Phys. Rev. E, \textbf{68}, 046117 (2003).

\bibitem{merton76}
R. C. Merton, J. Financ. Econ. \textbf{3}, 125 (1976).

\bibitem{gorenflo02}
R. Gorenflo, J. Loutchko, Yu Luchko, Fract. Calc. Appl. Anal. \textbf{5}, 491
(2002).

\bibitem{podlubny05}
I. Polubny and Martin Kacenak, mlf.m: \textit{Mittag-Leffler function ---
Calculates the Mittag-Leffler function with desired accuracy}, \textsc{MATLAB}
Central File Exchange, file {ID} \#8738 (2005),
www.mathworks.com/matlabcentral/fileexchange.

\bibitem{hilfer06}
R. Hilfer and H. J. Seybold, Integr. Transf. Spec. F., \textbf{17}, 637,
(2006).

\bibitem{samko93}
S. G. Samko, A. A. Kilbas, and O. Marichev, \textit{Fractional Integrals and
Derivatives, Theory and Applications} (Gordon and Breach Science Publishers,
London, 1993).

\bibitem{podlubny99}
I. Podlubny, \textit{Fractional Differential Equations} (Academic Press,
San Diego, 1999).

\bibitem{metzler99a}
R. Metzler, J. Klafter, and I. Sokolov, Phys. Rev. E \textbf{58}, 1621 (1999).

\bibitem{metzler99b}
R. Metzler, E. Barkai, and J. Klafter, Phys. Rev. Lett. \textbf{82}, 3563
(1999).

\bibitem{metzler99c}
R. Metzler, E. Barkai, and J. Klafter, Europhys. Lett. \textbf{46}, 431 (1999).

\bibitem{barkai00}
E. Barkai, R. Metzler, and J. Klafter, Phys. Rev. E \textbf{61}, 132 (2000).

\bibitem{magdziarz07}
M. Magdziarz, A. Weron, and K. Weron, Phys. Rev. E \textbf{75}, 016708 (2007).

\bibitem{vankampen81}
N. G. van Kampen, \textit{Stochastic Processes in Physics and Chemistry}
(North-Holland, Amsterdam, 1981).

\bibitem{risken92}
H. Risken, \textit{The Fokker-Planck Equation. Methods of Solution and
Applications}, 2nd edition with corrections (Springer, Berlin, 1992).

\bibitem{mainardi00}
F. Mainardi, M. Raberto, R. Gorenflo, and E. Scalas, Physica A \textbf{287},
468 (2000).

\bibitem{mura08}
A. Mura, M. S. Taqqu, and F. Mainardi, Physica A \textbf{387}, 5033 (2008).

\bibitem{gelfand64}
I. M. Gel'fand and G. E. Shilov, \textit{Generalized Functions} (Academic
Press, New York, 1964).

\bibitem{protter04}
P. Protter, \textit{Stochastic Integration and Differential Equations}, 2nd
edition (Springer, Berlin, 2004).

\bibitem{paul00}
W. Paul and J. Baschnagel, \textit{Stochastic Processes --- From Physics to
Finance} (Springer, Berlin, 2000).

\bibitem{gardiner85}
C. W. Gardiner, \textit{Handbook of Stochastic Methods}, 2nd edition,
(Springer, Berlin, 1996).

\bibitem{williams91}
D. Williams, \textit{Probability with Martingales} (Cambridge University Press,
Cambridge, UK, 1991).

\bibitem{schilling05}
R. L. Schilling, \textit{Measures, Integrals and Martingales} (Cambridge
University Press, Cambridge, UK, 2005).

\bibitem{devroye86}
L. Devroye, \textit{Non-Uniform Random Variate Generation} (Springer, New York,
1986).

\bibitem{devroye96}
L. Devroye, in \textit{Proceedings of the 1996 Winter Simulation Conference},
edited by J. M. Charnes, D. J. Morrice, D. T. Brunner, and J. J. Swain (IEEE
Press, New York, 1996), pp. 265--272.

\bibitem{chambers76}
J. M. Chambers, C. L. Mallows, and B. W. Stuck, J. Am. Stat. Assoc.
\textbf{71}, 340 (1999).

\bibitem{mcculloch96}
J. H. McCulloch, stabrnd.m: \textit{Stable random number generator},
\textsc{Matlab} script (1996), www.econ.ohio-state.edu/jhm/jhm.html.

\bibitem{kozubowski99}
T. J. Kozubowski and S. T. Rachev, Int. J. Comput. Numer. Anal. Appl.
\textbf{1}, 177 (1999).

\bibitem{germano08}
G. Germano, D. Fulger, and E. Scalas, mlrnd.m: \textit{Mittag-Leffler
pseudo-random number generator}, \textsc{Matlab} Central File Exchange,
file {ID} \#19392 (2008).



\bibitem{mainardi04}
F. Mainardi, R. Gorenflo and E. Scalas,
Vietnam J. Math. \textbf{32}, 53 (2004).

\bibitem{mainardi01}
F. Mainardi, Yu. Luchko and G. Pagnini,
Fract. Calc. Appl. Anal. \textbf{4}, 153 (2001).

\bibitem{meerschaert02}
M. M. Meerschaert, D. A. Benson, H.-P. Scheffler, P. Becker-Kern, Phys. Rev. E
\textbf{66}, 060102 (2002).

\bibitem{meerschaert06}
M. M. Meerschaert and E. Scalas, Physica A \textbf{370}, 114 (2006).

\bibitem{jacod03}
J. Jacod and A. N. Shiryaev, \textit{Limit Theorems for Stochastic Processes},
2nd edition, Vol. 288 of Grundlehren der mathematischen Wissenschaften
(Springer, Berlin, 2003).

\bibitem{press07}
W. H. Press, Saul A. Teukolsky, William T. Vetterling, Brian P. Flannery,
\textit{Numerical Recipes --- The Art of Scientific Computing}, 3rd edition
(Cambridge University Press, Cambridge, UK, 2007).

\end{thebibliography}
\end{document}